\documentclass[%
reprint,
twocolumn,
superscriptaddress,
showpacs,preprintnumbers,
nofootinbib,
aps,
prd,
]{revtex4-1}
\usepackage{graphicx,color}
\usepackage{amsmath,amssymb,amsfonts}
\usepackage{stackrel}
\usepackage{enumitem}
\usepackage[english]{babel}
\usepackage{verbatim}
\usepackage{hyperref}
\usepackage{comment}
\usepackage{ulem}
\usepackage{subcaption}
\usepackage{dcolumn}
\usepackage{physics}

\graphicspath{
  {./}
  {figures/}
}

\newcommand\he[1]{#1^\dagger} % Hermitian conjugate
\newcommand\gr[1]{\mathrm{#1}}% font for group names such as SU(2)

\newcommand\MSbar{$\overline{\text{MS}}$ } % MS-bar
\newcommand\Veff{V_\text{eff}} % Effective potential

\begin{document}

%% addresses

\newcommand{\HEL}{\affiliation{Department of Physics, PL 64, FI-00014
    University of Helsinki, Finland }}
    
\newcommand{\SJTU}{\affiliation{Tsung-Dao Lee Institute and School of
    Physics and Astronomy, Shanghai Jiao Tong University, 800
    Dongchuan Road, Shanghai, 200240 China }}

\newcommand{\UMASS}{\affiliation{Amherst Center for Fundamental
    Interactions, Department of Physics, University of Massachusetts,
    Amherst, MA 01003} }

\newcommand{\CIT}{\affiliation{Kellogg Radiation Laboratory,
    California Institute of Technology, Pasadena, CA 91125 USA} }

\newcommand{\BERN}{\affiliation{Albert Einstein Center for Fundamental
    Physics, Institute for Theoretical Physics, University of Bern,
    Sidlerstrasse 5, CH-3012 Bern, Switzerland}}

\newcommand{\Nottingham}{\affiliation{School of Physics and Astronomy,
    University of Nottingham, Nottingham NG7 2RD, UK}}

\title{Thermodynamics of a two-step electroweak phase transition}

\preprint{HIP-2020-11/TH; ACFI-T20-05}

\author{Lauri~Niemi}
\email{lauri.b.niemi@helsinki.fi}
\HEL

\author{Michael~J.~Ramsey-Musolf}
\email{mjrm@sjtu.edu.cn, mjrm@physics.umass.edu}
\SJTU\UMASS\CIT

\author{Tuomas~V.~I.~Tenkanen}
\email{tenkanen@itp.unibe.ch}
\BERN

\author{David~J.~Weir}
\email{david.weir@helsinki.fi}
\HEL
\Nottingham

\begin{abstract}
New field content beyond that of the Standard Model of particle
physics can alter the thermal history of electroweak symmetry breaking in the
early universe. In particular, the symmetry breaking may have occurred
through a sequence of successive phase transitions. We study the
thermodynamics of such scenario in a real triplet extension of the
Standard Model, using nonperturbative lattice simulations. Two-step
electroweak phase transition is found to occur in a narrow region of
allowed parameter space with the second transition always being first
order. The first transition into the phase of non-vanishing triplet
vacuum expectation value is first order in a non-negligible portion of the two-step parameter space. A
comparison with 2-loop perturbative calculation is provided and
significant discrepancies with the nonperturbative results are identified.
\end{abstract}

\maketitle

\section{Introduction}

In the Standard Model (SM) of particle physics, electroweak (EW) gauge
symmetry is spontaneously broken by the vacuum-expectation value (VEV)
of the Higgs field. Thermal corrections to the Higgs potential restore
this symmetry in the early universe. For the physical Higgs mass this
transition is a smooth crossover rather than a true phase transition
\cite{Kajantie:1996mn,Csikor:1998eu,DOnofrio:2015gop}, \textit{i.e.},
there is no distinction between the symmetric and broken
``phases''. In many beyond the Standard Model (BSM) scenarios, the
introduction of additional scalar fields can result in a scalar
potential having vastly different thermal behavior from that of the
SM. In particular, these extensions may yield a {\it bona fide}
electroweak phase transition (EWPT) that is first order, with
cosmological consequences that include conditions needed to generate
the cosmic matter-antimatter asymmetry through electroweak
baryogenesis (EWBG)
\cite{Kuzmin:1985mm,Shaposhnikov:1986jp,Shaposhnikov:1987tw} and
production of gravitational waves (GW). A conclusive test of this
possibility could result from present and future high energy collider
experiments \cite{Ramsey-Musolf:2019lsf} and GW
probes~\cite{Caprini:2015zlo,Huang:2016odd,Caprini:2019egz}.

An extended scalar potential may admit a richer thermal history than
in the SM. The new fields may have phase transitions of their own, and
the universe may undergo several symmetry-breaking transitions before
settling down to the present EW vacuum. While such a thermal history
would be interesting in itself, multi-step EW symmetry breaking could
have important implications for cosmology.  Specifically, EWBG could
be realized in a sequence of symmetry-breaking transitions around the
EW scale \cite{Hammerschmitt:1994fn,Patel:2012pi,Inoue:2015pza,Blinov:2015sna,Ramsey-Musolf:2017tgh}. This setup
also leads naturally to a strong first-order transition into the final
EW phase through a tree-level potential barrier. Furthermore, a
non-minimal pattern of EW symmetry breaking can produce topological
solitons, such as monopoles and domain walls, with potentially
interesting properties. Such defects are absent in the SM, but are
generic in grand unified theories~\cite{Jeannerot:2003qv}; many
analogues also exist in condensed matter systems~\cite{Zurek:1996sj}.

The simplest extension of the SM scalar sector admitting distinct
phases of broken EW symmetry in the early universe is the real triplet
model with three BSM degrees of freedom, collectively denoted by
$\Sigma$.  In the resulting ``$\Sigma$SM'', EW symmetry breaking
may occur directly in a single step from the unbroken phase $O$ to the
Higgs phase $\phi$, or in two steps, $O\to\Sigma\to\phi$, where EW
symmetry is broken in both the $\Sigma$ and $\phi$ phases. A
delineation of the model parameters leading to either possibility is
given in the perturbative analysis in \cite{Patel:2012pi}. Analogous
studies in other models containing new scalars either charged or
neutral under the SM gauge symmetries indicate that multi-step
transitions may arise generically \cite{Hammerschmitt:1994fn,Profumo:2007wc,Espinosa:2011ax,Patel:2013zla,Curtin:2014jma,Blinov:2015sna,Jiang:2015cwa,Kurup:2017dzf,Chiang:2017nmu,Kang:2017mkl}. Thus, a more thorough
investigation of the thermal history and phase diagram of the
$\Sigma$SM is well-motivated.

A robust determination of the phase diagram is a nontrivial task even
for theories that are weakly coupled at zero temperature. The EWPT is
driven by infrared (IR) bosonic fields, the Matsubara zero modes,
whose mutual interactions are boosted by Bose enhancement. This
results in a poor convergence of perturbation theory and ultimately
renders the momentum scale $\sim g^2 T$ nonperturbative, $g$ being a
gauge coupling \cite{Linde:1980ts}. This problem affects gauge bosons
in the symmetric high-temperature phase and scalar fields near a phase
transition where their correlation lengths can grow large. Indeed,
perturbation theory incorrectly predicts a first-order EWPT in the
minimal SM. There is no \textit{a priori} reason to trust the
perturbative description in BSM settings either, unless one is
interested solely in properties of the Higgs phase, where the VEV
provides a perturbative mass for most excitations. Large couplings in
the scalar sector may further aggravate the IR problem
\cite{Laine:2017hdk,Kainulainen:2019kyp}.

For the EW theory, a solution to the IR problem is known: the
thermodynamics are well described by a three-dimensional (3d)
effective field theory (EFT) for which nonperturbative lattice
simulations can be carried out
\cite{Ginsparg:1980ef,Appelquist:1981vg,Kajantie:1995dw}. This
``dimensional reduction'' amounts to perturbatively integrating out
nonzero Matsubara modes, and the resulting theory describes thermal
fluctuations of the bosonic zero modes.

Here, we report on a nonperturbative study of the $\Sigma$SM using the
3d EFT. The results are used to obtain a realistic picture of the
two-step EWPT scenario. We also assess the performance of the
perturbative treatment in light of our nonperturbative results.

\section{Model}
The color neutral scalar field $\Sigma$ carries no
hypercharge, transforms under the adjoint representation of
$\gr{SU(2)}_L$, and does not couple to SM fermions.  For simplicity,
we further require invariance under the $Z_2$ transformation $\Sigma
\rightarrow -\Sigma$, which allows for the VEV $v_\Sigma$ to vanish at
$T=0$. Doing so ensures consistency with bounds on the EW $\rho$
parameter while enabling the neutral field $\Sigma^0$ to contribute to
the dark matter relic density
\cite{FileviezPerez:2008bj,Cirelli:2005uq}. Recent studies of the
corresponding collider and dark matter phenomenology appear in
\cite{Bell:2020gug,Chiang:2020rcv}. The most general, renormalizable
scalar potential then reads
\begin{align}
V(\phi,\Sigma) =& -\mu_\phi^2 \he\phi\phi - 
\frac12 \mu_\Sigma^2 \Sigma^a \Sigma^a 
+ \lambda (\he\phi\phi)^2 
\nonumber \\ & 
+\frac{b_4}{4} 
(\Sigma^a\Sigma^a)^2 
+ \frac{a_2}{2} \he\phi\phi 
\Sigma^a \Sigma^a,
\end{align}
where $a=1,2,3$ is the adjoint index, with $\sqrt{2}
\Sigma^\pm=\Sigma^1\mp i\Sigma^2$ and $\Sigma^0=\Sigma^3$.

For $\mu_\phi^2 > 0$, the potential has a symmetry-breaking minimum in
the Higgs direction, $\langle \he\phi\phi\rangle = \frac12 v^2$ with
$v_\Sigma=0$.  This corresponds to the standard EW minimum with three
BSM excitations from the $\Sigma$ field, whose masses are degenerate
at tree level~\cite{FileviezPerez:2008bj}. Following
\cite{Niemi:2018asa}, we relate the Lagrangian parameters to EW
observables through pole-mass renormalization at one-loop level,
taking the mass $M_\Sigma$ of $\Sigma^0$ as an input parameter. We
treat the couplings $a_2$ and $b_4$ as input parameters directly at
the \MSbar scale $M_Z$. The one-loop correction is necessary to match
the accuracy of our EFT construction below.

If $\mu_\Sigma^2 > 0$, a second minimum of $V(\phi,\Sigma)$ appears in
the $\Sigma$ direction, with $v=0$. Physics in this $\Sigma$ vacuum
resembles that of the broken phase of the $\gr{SU(2)}$ Georgi-Glashow
model~\cite{Georgi:1972cj}: $\gr{SU(2)}$ breaks to a $\gr{U(1)}$ gauge group
distinct from that of the usual electromagnetic
interaction.

In the $\Sigma$ vacuum phase, the system admits 't Hooft-Polyakov
monopole
excitations~\cite{Balian:2005joa,tHooft:1974kcl,Polyakov:1974ek}. These
are topological soliton solutions of the field equations, carrying a
magnetic charge under the residual $\mathrm{U}(1)$ gauge group. When
the system crosses to the $\Sigma$ vacuum, these monopoles can
freeze-out as a result of existing long-wavelength thermal
fluctuations~\cite{Rajantie:2002dw}. They may also play a role in the
dynamics of the finite-temperature phase transition
itself~\cite{Kajantie:1997tt}.

Thermal corrections can modify the vacuum structure. In the
high-$T$ limit, the leading effect is a $T$-dependent reduction of the
squared mass parameters: $\mu_{\phi,\Sigma}^2 \rightarrow
\mu_{\phi,\Sigma}^2 - \Pi_{\phi,\Sigma} T^2$.  Here
$\Pi_{\phi,\Sigma}$ are $\mathcal{O}(g^2)$ constants, where for
notational convenience $g^2$ denotes a general quartic coupling. The
thermal correction turns $\mu^2_\phi$ negative at $T_\phi \sim 100$
GeV, relaxing the Higgs VEV to zero. Two-step EWSB occurs if the
thermal corrections drive $\mu_{\Sigma}^2$ negative at a higher
temperature $T_\Sigma > T_\phi$. The universe then resides in the
symmetric phase $O$ at high temperatures before transitioning into the
$\Sigma$ phase ($O \rightarrow \Sigma$) at $T_\Sigma$, followed by
another phase transition ($\Sigma\rightarrow \phi$) into the final
Higgs phase at $T_\phi$. The presence of a tree-level saddle point
separating the $\phi$ and $\Sigma$ minima suggests a first-order
transition in the second stage.

\section{Effective theory at high temperature}

We derive the 3d EFT in the imaginary time
formalism by integrating out modes with a nonzero Matsubara frequency,
including all fermions, leading to the Euclidean space Lagrangian:
\begin{align}
\label{eq:3d-EFT}
\mathcal{L}_\text{3d} =& \frac14 (F^a_{ij})^2 + |D_i \phi|^2 + \frac12 (D_i \Sigma^a)^2 + \bar{\mu}_\phi^2 \he\phi\phi + \bar{\lambda} (\he\phi\phi)^2 \nonumber \\ & + \frac{\bar{\mu}_\Sigma^2}{2} \Sigma^a\Sigma^a  + \frac{\bar{b}_4}{4} (\Sigma^a\Sigma^a)^2 + \frac{\bar{a}_2}{2} \he\phi\phi \Sigma^a \Sigma^a.
\end{align}
Here $F^a_{ij}$ is the $\gr{SU(2)}_L$ field strength tensor. Thermal
corrections from the hard scale $\pi T$ are included in the barred
parameters, whose matching was worked out to $\mathcal{O}(g^4)$
accuracy in \cite{Niemi:2018asa} and includes corrections from
temporal components of the gauge fields that generate a Debye
screening mass and can be integrated out
\cite{Kajantie:1995dw,Niemi:2018asa}. These couplings in
(\ref{eq:3d-EFT}) are dimensionful, and the fields are scaled by
$T^{-1/2}$. We have neglected the $\gr{U(1)}_Y$ gauge field and the
$\gr{SU(3)}_C$ sector as they have only a small effect on the EWPT
\cite{Kajantie:1996qd} and do not couple to $\Sigma$.

The EFT is formally valid in the high-$T$ limit $m \ll \pi T$. By
construction, its region of validity overlaps with that of the
consistent daisy resummation of \cite{Arnold:1992rz} as required to
correctly describe physics at the ``soft'' scale $gT$. The EFT
systematically includes these corrections.

\begin{figure}[t]
    \centering
    \includegraphics[width=0.5\textwidth]{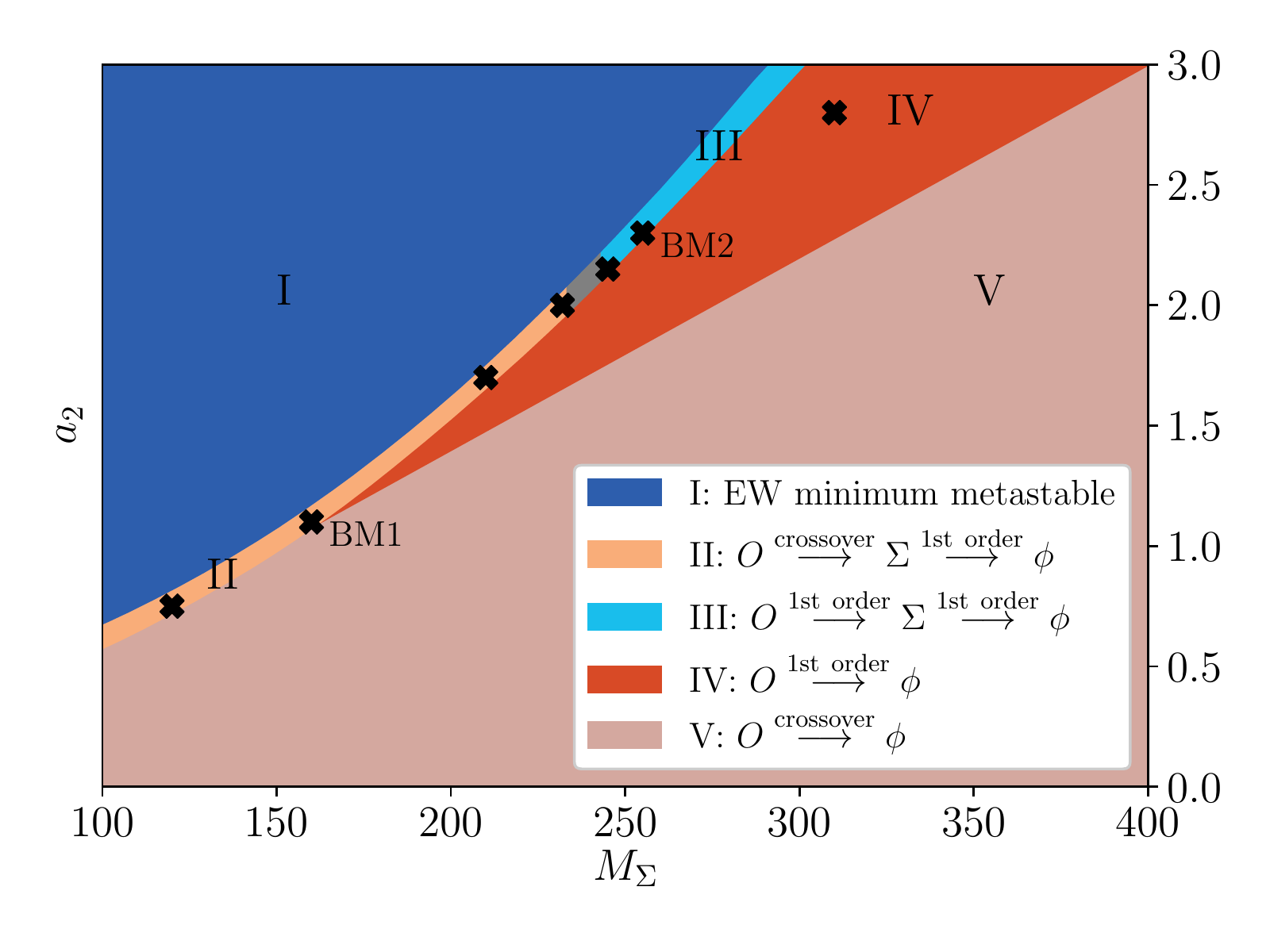}
    \caption{\label{fig:phasediagram} $\Sigma$SM phase diagram for $\Sigma$ self-coupling parameter $b_4=0.25$. Vertical and horizontal axes give the triplet-Higgs coupling and triplet mass, respectively. 
Colored regions correspond to different types of EW symmetry-breaking transitions. The allowed parameter space is
      dominated by direct transitions into the Higgs phase (regions IV
      and V). Regions II and III lead to a two-step symmetry-breaking history
      distinguished by whether the $O\rightarrow \Sigma$ transition is a
      crossover; a first order EWPT; or a second-order EWPT (a line somewhere in the
      grey region). Both $O \rightarrow \phi$ and $O \rightarrow
      \Sigma$ transitions grow stronger as the quartic portal coupling
      $a_2$ increases. In region I the EW minimum is not the global
      minimum at zero temperature, according to the one-loop, $T=0$ effective
      potential. Our lattice benchmarks are marked with a
      cross; points BM1 and BM2 are discussed in detail below.  }
\end{figure}

To probe the parameter space for a two-step EWPT, we have scanned the
parameters using the effective potential $\Veff$ calculated to
two-loop order in the EFT. Evolution of the different minima is
tracked using the gauge-invariant approach described in
\cite{Laine:1994zq,Patel:2011th}; details of the calculation can be
found in the Appendix. Two-step transitions occur in a narrow band separating the parameter
space of one-step EWPTs ($O \rightarrow \phi$) from that where the EW
minimum is metastable at $T=0$. For $b_4 = 0.25$, this is illustrated
in Fig.~\ref{fig:phasediagram}.

In a vast region of parameter space, the EWPT is driven solely by the
Higgs doublet, which becomes parametrically light near the critical
temperature, $\bar{\mu}^2_\phi \sim (g^2T)^2$, due to a cancellation
between vacuum and thermal masses. This allows us to integrate out
$\Sigma$ as an UV mode near the critical temperature $T_c$, resulting
in a simpler EFT for which the nonperturbative phase diagram is known
\cite{Kajantie:1995kf,Kajantie:1996mn}. This approach was taken in
\cite{Niemi:2018asa} to identify where the $O \rightarrow \phi$
transition is a crossover: region V of
Fig.~\ref{fig:phasediagram}. Deep in region IV, integrating out
$\Sigma$ is no longer justified, but our simulations verify that the
transition remains first order here. The line separating regions IV
and V corresponds to second order transitions. Its location is only
accurate within $\sim 10\%$ due to neglect of higher-dimensional
operators \cite{Niemi:2018asa}.

It is interesting to ask where in the two-step EWPT region
the $O \rightarrow \Sigma$ transition is first order. As in the SM
case, perturbation theory does not provide reliable guidance; genuine
nonperturbative input is needed. Qualitatively, at temperatures near a
$O \rightarrow \Sigma$ transition we expect the IR physics to resemble
that of a Georgi-Glashow type theory containing just gauge fields and
$\Sigma$. The corresponding phase transition terminates at a finite
value of the scalar self coupling \cite{Kajantie:1997tt}. Our
simulations confirm this expectation: the $O \rightarrow \Sigma$
transition is crossover in region II; first order in region III;
and terminates somewhere in the grey region between. We have not attempted a more
precise determination of the end line.
 
The IR behavior also suggests that the $O \rightarrow \Sigma$
transition grows stronger at small $b_4$, which we have verified 
with simulations using $b_4 = 0.15, 0.20$. However, the two-step
region itself becomes narrower due to a decrease in the $\Sigma$
minimum vacuum energy, which goes as $\sim -\mu^4_\Sigma / (4b_4)$ at
tree level. There is no two-step EWPT if the $T=0$ potential is deeper
in the $\Sigma$ direction than in the Higgs minimum (region I).

\section{Simulations}

Simulations in the full $\Sigma$SM are not
practical due to the chirally coupled fermions. A systematic method
for implementing the fermionic corrections (which are significant) is
provided by the dimensionally-reduced EFT (\ref{eq:3d-EFT}). To
discretize it, we employ the (unimproved) Wilson action for the gauge
links and couple these to the scalars through gauge-invariant hopping
terms. Parameters in the lattice action
are related to the continuum parameters in Eq.~(\ref{eq:3d-EFT}) by
expressions given in \cite{Laine:1995np}. These relations become exact
in the continuum limit as a consequence of super-renormalizability of
the 3d EFT.

In lattice simulations, we determine probability distributions of
gauge-invariant operators by generating field configurations in the
canonical ensemble. For the EWPT, the observables of interest are
scalar condensates, particularly $\langle \phi^\dagger\phi\rangle$ and
$\langle \Sigma^a \Sigma^a \rangle$, whose probability distributions
in a first order transition develop a two-peak structure. The peaks
correspond to the bulk phases and have equal integrated probabilities
at $T_c$ \cite{Kajantie:1995kf}.

\begin{figure*}[t!]
    \centering
    \begin{subfigure}[b]{0.50\textwidth}
        \centering
        \includegraphics[width=\textwidth]{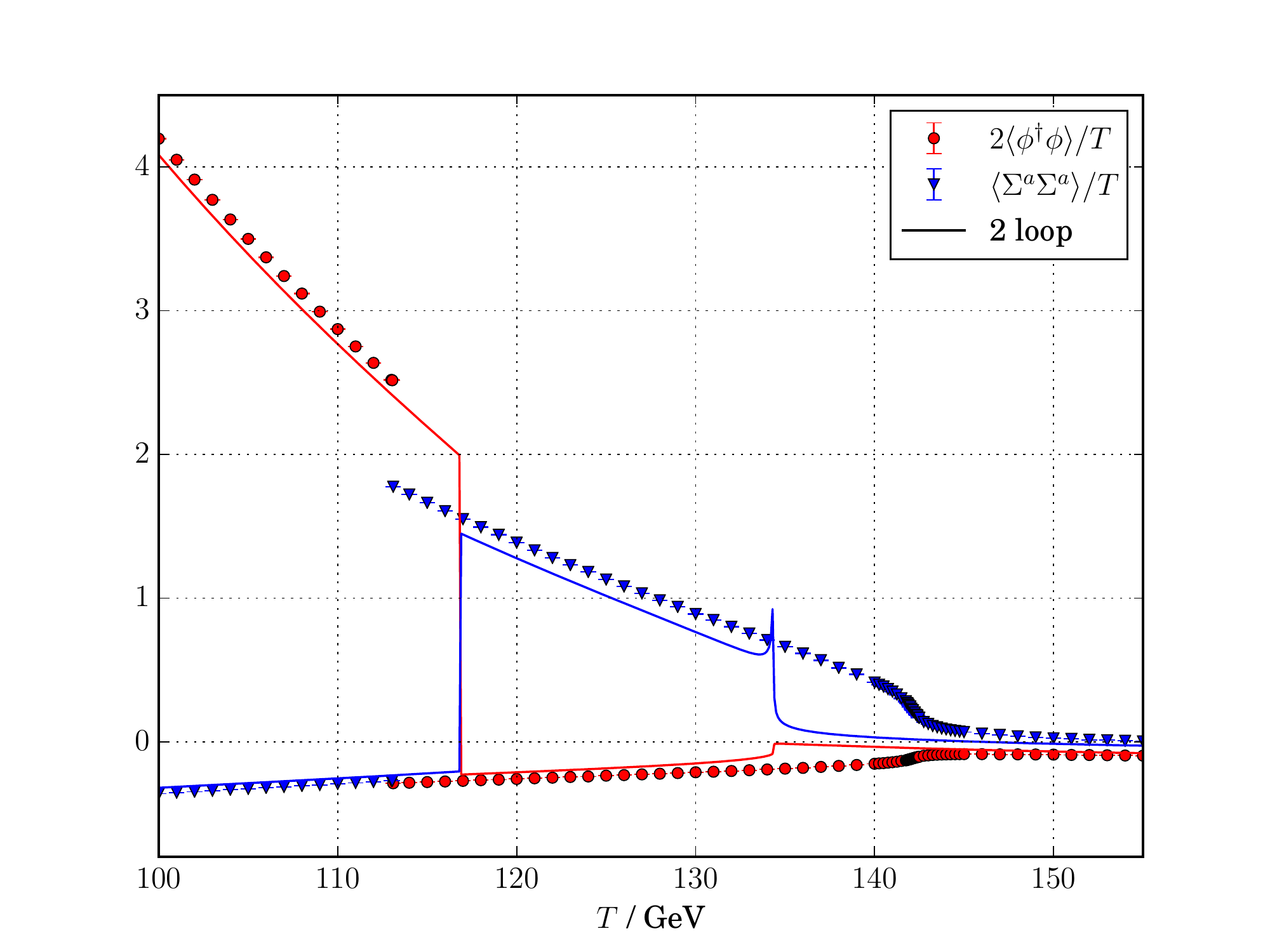}
        \caption{\centering{BM1: $(M_\Sigma, a_2, b_4)$
            = $(160\text{ GeV}, 1.1, 0.25)$}}
    \end{subfigure}%
    ~ 
    \begin{subfigure}[b]{0.50\textwidth}
        \centering
        \includegraphics[width=\textwidth]{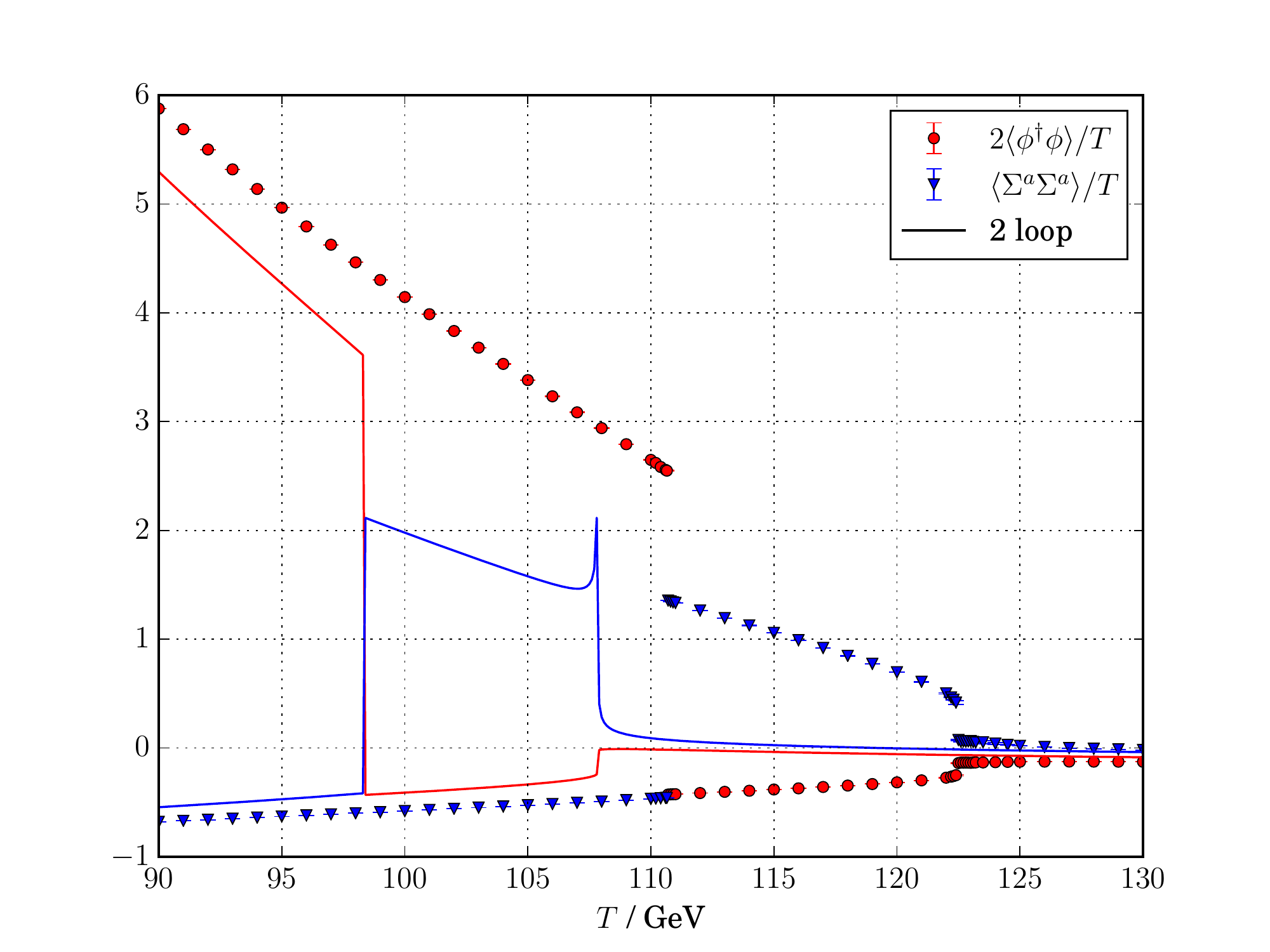}
        \caption{\centering{BM2: $(M_\Sigma, a_2, b_4)$
            = $(255\text{ GeV}, 2.3, 0.25)$}}
    \end{subfigure}
    \caption{\label{fig:condensates} Temperature-dependence of quadratic condensates in the 3d EFT, as measured on the lattice in the global
      probability maximum and converted to the \MSbar scheme (at scale
      $T$) using relations  in Ref.~\cite{Laine:1995np}. The
      solid lines are perturbative estimates at two loop order in the same EFT. Results shown are for a $60^3$ lattice with 
      $4/(a\bar{g}^2) = 24$, large enough for finite-size
      effects to be negligible. Monte Carlo statistical errors
      are too small to be visible at this scale.}
\end{figure*}

In the region separating the bulk phases, the ensemble is dominated by
mixed-phase configurations where the two phases exist simultaneously
on the lattice \cite{Moore:2000jw}. The phase interface carries free
energy proportional to its surface area, and the probability of
tunneling between phases is thus exponentially suppressed. On large
lattices, this makes it difficult to obtain the probability
distributions using conventional update algorithms for the canonical
ensemble. Instead, we apply multicanonical simulations to boost the
probabilities of the mixed configurations relative to the bulk phases
\cite{Berg:1991cf}. The ensemble is modified by a suitable weight
function $W$ as $\exp[-S] \rightarrow \exp[-S -
  W(\Phi_\text{multi})]$, where $\Phi_\text{multi}$ is typically an
order parameter-like quantity that distinguishes the phases. The
canonical distributions are then obtained by reweighting the
measurements \cite{Ferrenberg:1988yz}. While $W$ itself can be
calculated recursively \cite{Laine:1998qk,Wang:2000fzi}, the
efficiency of multicanonical simulations depends on the choice of
$\Phi_\text{multi}$. For $O \rightarrow \Sigma$ transitions, we choose
the volume average of $\Sigma^a\Sigma^a$, and the simulations proceed
analogously to those of Refs.~\cite{Kajantie:1995kf,Kajantie:1997tt}.

Consistent with perturbation theory we found that, for all the cases
we studied (crosses in Fig.~\ref{fig:phasediagram}), the $\Sigma \rightarrow \phi$ stage is a first-order
transition with strong suppression of the mixed configurations. We
have not found a simple choice of $\Phi_\text{multi}$ that would
efficiently take the system both ways between the two broken
phases. In either phase, one of the scalar condensates develops large
bulk fluctuations, and an even larger fluctuation is required to start
the tunneling process. Instead, we determine $T_c$ by restricting the
simulation to sample the mixed-phase configurations only. At $T_c$,
neither phase is preferred over the other, and probability
distributions of order parameters in the allowed range become
approximately flat.

The simulation results carry mild dependence on lattice volume and
spacing $a$, but extrapolations $V\rightarrow \infty$ and $a\rightarrow 0$
can be taken in a controlled fashion
\cite{Kajantie:1995kf,Laine:2000rm,Laine:2012jy}. For the first-order
transitions studied here, $\mathcal{O}(a)$ errors appear negligible
for $4/(a\bar{g}^2) \geq 20$, with both $T_c$ and condensate values changing
by less than $1\%$ if $a$ is decreased. Volume dependence appears to
be even smaller, suggesting that our finite-size effects are well
under control. Below we quote results from only the largest lattices.

Our code for simulating the $\gr{SU(2)}$ theory with fundamental and
adjoint Higgses has been cross-checked by reproducing histograms in
Refs.~\cite{Kajantie:1995kf,Kajantie:1997tt}. We employ conventional
heatbath updates for the gauge links \cite{Kennedy:1985nu} and a
mixture of Metropolis and over-relaxation updates
\cite{Kajantie:1995kf} for the scalars.

\section{Results \& Discussion}

The condensates require additive
renormalization, but their discontinuities across a phase transition
are renormalization group invariant and directly related to the latent
heat $L$ \cite{Farakos:1994xh,Laine:1995np}, a physical quantity
characterizing transition strength. Fig.~\ref{fig:condensates} shows
the condensate evolution for two benchmark (BM) points giving a
two-step EWPT, together with perturbative estimates. In the high-$T$
phase the condensates stay close to zero, while at low temperatures
$\langle \he\phi\phi\rangle$ obtains a large value. The existence of
an intermediate $\Sigma$ phase is clearly visible. The condensates can
be negative because of the additive renormalization.

\begin{figure}
	\begin{center}
		\includegraphics[width=0.50\textwidth]{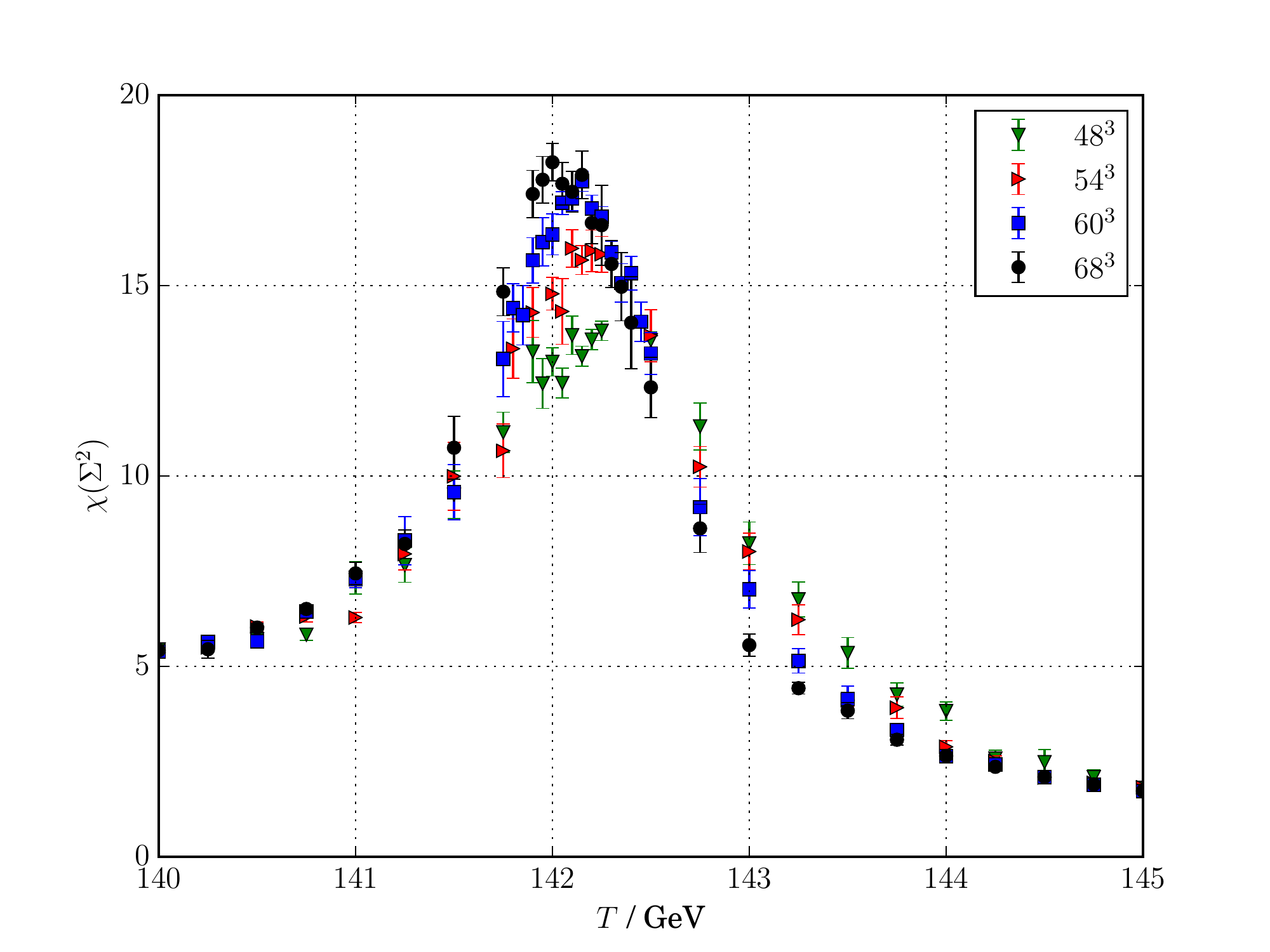}
	\end{center}
	\caption{$T$-dependence of $\Sigma^a \Sigma^a$-susceptibility (\ref{eq:suscept}) across the
          crossover in BM1, measured at $4/(a\bar{g}^2) = 24$ and different
          volumes.}
	\label{fig:susc}
\end{figure}

In BM1, the $O \rightarrow \Sigma$ stage is a crossover: we find no
evidence of phase coexistence, ruling out a first-order transition. To
investigate the possibility of a second order transition we studied
finite-size scaling of the dimensionless $\Sigma^2$ susceptibility,
\begin{align}
  \chi(\Sigma^2) = \frac14 V T \left[  \left\langle (\Sigma^a\Sigma^a)_V^2
    \right\rangle - \left\langle (\Sigma^a\Sigma^a)_V \right\rangle^2 \right]
    \label{eq:suscept}
\end{align}
where the subscript denotes volume averaging. As shown in
Fig.~\ref{fig:susc}, $\chi(\Sigma^2)$ peaks at $T\approx 142$ GeV but
converges to a finite value as $V\rightarrow \infty$, consistent with
crossover behavior. By contrast, for a second-order transition the
susceptibility diverges with a critical exponent. The first transition
in BM2 is first order and, depending on the criterion for baryon
number preservation within the $\Sigma$ phase, could be strong enough
to support two-step EWBG \cite{Patel:2012pi}.

To assess the reliability of perturbation theory, we compare the
nonperturbative results to those obtained from the two-loop $\Veff$
(solid lines in Fig.~\ref{fig:condensates}). In the gauge-invariant
treatment used here, the potential is minimized by expanding the VEVs
around their tree-level values (see the Appendix). Near the $O
\rightarrow \Sigma$ transition, this approach breaks down due to the
absence of a small expansion parameter, and the potential encounters
an IR divergence \cite{Laine:1994zq,Farakos:1994xh}. This is the
reason for the spiking of $\langle \Sigma^a\Sigma^a \rangle / T$ in
fig.~\ref{fig:condensates}. Consequently, the crossover in BM1 is not
visible perturbatively.

Outside the temperature range of $O\rightarrow \Sigma$ transitions,
perturbation theory provides some rough qualitative guidance but
performs poorly in quantitatively describing both $T_c$ and the
``strength'' (condensate discontinuities). For the $\Sigma \rightarrow
\phi$ transition this finding is, perhaps, surprising, 
as the two minima are present already in the tree-level potential. Nevertheless, scalar loops can significantly alter the transition dynamics because of the large $a_2$ coupling necessary for a two-step EWPT. On dimensional grounds, the high-$T$ expansion parameter for $\phi-\Sigma$ interactions is of the form 
$a_2 T \times (\text{scalar mass})^{-1}$. At the second transition, the masses are bounded from below by the non-zero VEVs but are still numerically small compared to $a_2 T$ on both sides of the transition. Hence, a low-order perturbative description is not necessarily reliable. There may also be additional nonperturbative effects from magnetic monopoles that exist in the $\Sigma$ phase.

After extrapolating $V\rightarrow \infty$ and $a \rightarrow 0$, the
latent heat is $L/T_c^4=0.4109(2)$ for the $\Sigma \rightarrow \phi$
transition in BM1; in BM2 the value for the first (second) transition
is $0.151(2)$ ($0.5895(9)$). Errors are the stastistical
uncertainties. The perturbative values, where applicable, are smaller
by $30\%$ in BM1 and larger by $40\%$ in BM2. The discrepancy is
dominated by the error in $T_c$. The two-loop potential is crucial for
even a qualitative agreement with the nonperturbative results: at one
loop, the jumps in condensates are more than $50\%$ smaller than at
two loops, while the temperatures differ only by a few percent.

Applicability of these results to the full 4d $\Sigma$SM depends on
the overall accuracy of our 3d EFT.  Dimensional reduction produces
operators of dimension six (in 4d units) that we have neglected
here. We anticipate that the operators $c_\phi (\he\phi\phi)^3 / T^2$
and $c_\Sigma (\Sigma^a\Sigma^a)^3 / T^2$ yield the largest
contribution, with a potentially significant effect in the presence of
a non-vanishing condensate. Following \cite{Kajantie:1995dw}, we
estimate their effects on scalar VEVs at tree level. For $T > 50$ GeV,
the operators cause relative shifts of less than $1\%$ in the VEVs in
both BM1 and BM2, suggesting that the performance of our dimensional
reduction is comparable to the SM case, despite the relatively heavy
scalar excitations in the Higgs phase; indeed,  top
quark contributions still dominate $c_\phi$.

Overall, our results for the $\Sigma$SM phase diagram
(Fig.~\ref{fig:phasediagram}) validate the expectations from purely
perturbative studies that the early universe could have undergone
successive EWSB transitions. To our knowledge, this work provides the
first non-perturbative demonstration of this possiblility. At the same
time, a robust determination of the character of these transitions and
a quantitative determination of their properties ($T_C$, latent heat,
and model parameter-dependence) requires a nonperturbative
treatment. Looking ahead, we anticipate that future nonperturbative
studies will be essential for obtaining dynamical properties ({\it
  e.g.}, rates for nucleation \cite{Moore:2000jw}, sphaleron
transitions \cite{DOnofrio:2014rug}, and monopole-catalyzed processes
\cite{Rubakov:1981rg}) necessary for a complete picture of the associated
thermal history in the $\Sigma$SM and other extended scalar sector
scenarios. In this context, we consider the present study as the first
step in a exciting program aimed at building a rigorous understanding
of non-minimal electroweak symmetry breaking.

\begin{acknowledgments}
We thank Oliver Gould, Mark Hindmarsh, Kimmo Kainulainen, Mikko Laine,
Arttu Rajantie and Kari Rummukainen for discussions, as well as
Philipp Schicho and Juuso {\"O}sterman for useful correspondence on 3d
loop integrals. LN acknowledges financial support from the Jenny and
Antti Wihuri Foundation. This work was partly supported by the Swiss
National Science Foundation (SNF) under grant 200020B-188712, and by
Academy of Finland under grants 308791 and 320123. MJRM was supported
in part under U.S. Department of Energy contract No. DE-SC0011095 and
National Science Foundation of China grant No. 19Z103010239. We are
grateful for computational resources provided by the University of
Helsinki clusters (urn:nbn:fi:research-infras-2016072533). DJW (ORCID
ID 0000-0001-6986-0517) was supported an Science and Technology
Facilities Council Ernest Rutherford Fellowship, grant
no. ST/R003904/1, by the Academy of Finland, grants 324882 and 328958,
and by the Research Funds of the University of Helsinki.
\end{acknowledgments}

\begin{widetext}

\appendix
\allowdisplaybreaks
\begin{center}{\large \textbf{Appendix}} \end{center}

\section{Gauge-invariant effective potential to two loops}

Here we collect details of the perturbative calculation that was used for comparison with the nonperturbative results in the main text. The goal is to compute thermal corrections to the effective potential $\Veff$ and extract from it values for $T_c$, latent heat and the condensates, and we do this at two-loop level. The perturbative expansion of $\Veff$ in terms of quartic couplings has a peculiar structure at finite temperature, with fractional powers such as $\lambda^{3/2}$ appearing as a consequence of Debye screening. A consistent inclusion of these effects requires daisy resummation in the high-$T$ approximation \cite{Arnold:1992rz}, but as discussed in the main text, it is easier to work directly in the 3d EFT given in Eq.~(\ref{eq:3d-EFT}), where these resummations are incorporated automatically. We take this approach, generalizing the calculation of \cite{Farakos:1994kx} to include a background field for the triplet. 

Parameters of the EFT are related to those in the full theory by matching relations presented in \cite{Niemi:2018asa}. Here we simplify the notation by dropping the overline from the EFT parameters. In what follows, all parameters are assumed to be those of the 3d theory (\ref{eq:3d-EFT}) and therefore temperature dependent. For completeness we also include the $U(1)_Y$ hypercharge field $B_i$, so the covariant derivatives read
\begin{align}
D_i \phi = (\partial_i - \frac12 i g \sigma_a A^a_i - \frac12 i g' B_i) \phi, \quad 
D_i \Sigma^a = \left(\partial_i \Sigma^a + i g \epsilon^{abc} A^b_i \Sigma^c\right).
\end{align}
For comparison with the nonperturbative results we have set $g' = 0$, as the $U(1)_Y$  field is not included in our lattice simulations.

We parametrize the scalars as 
\begin{align}
\phi = \frac{1}{\sqrt{2}}
\begin{pmatrix}
\phi_1 + i \phi_2 \\
v + \phi_3 + i \phi_4
\end{pmatrix} ,
\quad 
\vec{\Sigma} = 
\begin{pmatrix}
\Sigma_1 \\
\Sigma_2 \\
x + \Sigma_3
\end{pmatrix},
\end{align}
where $v$ and $x$ are real background fields. The Euclidean Lagrangian (\ref{eq:3d-EFT}) becomes
\begin{align}
\mathcal{L}_\text{3d} =& V_\text{tree}(v, x) + \mathcal{L}^{(2)}_\text{3d} + \mathcal{L}^{(I)}_\text{3d},  \\
\label{eq:3d-Veff-tree}
V_\text{tree}(v, x) =& \frac12 \mu_\phi^2 v^2 + \frac12 \mu_\Sigma^2 x^2 + \frac14 \lambda v^4 + \frac14 b_4 x^4 + \frac14 a_2 v^2 x^2.
\end{align}
Here $\mathcal{L}^{(2)}_\text{3d}$ and $\mathcal{L}^{(I)}_\text{3d}$ contain quadratic and interaction terms respectively. Terms linear in $\phi_i$ or $\Sigma_i$ do not contribute to the effective potential, which is defined (at a finite volume $\mathcal{V}$) through
\begin{align}
\label{eq:Veff-definition}
\exp\left[ -\frac{\mathcal{V}}{\hbar} \Veff(v,x)  \right] =& \int D\phi \; \exp\left[ -\frac{S_\text{3d}}{\hbar}\right] \nonumber \\
=& \exp\left[ -\frac{\mathcal{V}}{\hbar} V_\text{tree}(v,x) \right] \int D\phi \; \exp\left[ -\frac{1}{\hbar} \int d^3x \; \mathcal{L}^{(2)}_\text{3d} \right] \left\langle \exp\left[ -\frac{1}{\hbar} \int d^3x \; \mathcal{L}^{(I)}_\text{3d} \right]\right\rangle.
\end{align}
The symbolic measure $D\phi$ denotes functional integration over all dynamical fields, and the expectation value is to be calculated perturbatively.

As discussed in \cite{Patel:2011th}, the value of $\Veff$ in its minimum is guaranteed, by Nielsen identities, to be gauge invariant order-by-order in the loop-counting parameter $\hbar$. Expanding the potential and its minima as 
\begin{align}
\label{eq:exp-Veff-min}
\Veff = V_0 + \hbar V_1 + \hbar^2 V_2, \quad
v_\text{min} = v_0 + \hbar v_1 + \hbar^2 v_2, \quad
x_\text{min} = x_0 + \hbar x_1 + \hbar^2 x_2
\end{align}
and generalizing the analysis of \cite{Laine:1994zq,Patel:2011th} to the case of two background fields gives the \lq\lq $\hbar$ expansion"
\begin{align}
\Veff(v_\text{min}, x_\text{min}) =& V_0(v_0, x_0) + \hbar V_1(v_0, x_0) + \hbar^2 \left[V_2(v_0, x_0) - \frac12 v_1^2 \pdv[2]{V_0}{v} - \frac12 x_1^2 \pdv[2]{V_0}{x} - v_1 x_1 \pdv{V_0}{v}{x} \right] + \mathcal{O}(\hbar^3), \\
\label{eq:min-corrections}
v_1 =& \left[\left( \pdv{V_0}{v}{x} \right)^2 - \left(\pdv[2]{V_0}{v}\right) \left(\pdv[2]{V_0}{x} \right) \right]^{-1} \left[ \left( \pdv[2]{V_0}{x} \right)\left( \pdv{V_1}{v} \right) - \left( \pdv{V_0}{v}{x} \right)\left( \pdv{V_1}{x} \right) \right], \\
x_1 =& \left[\left( \pdv{V_0}{v}{x} \right)^2 - \left(\pdv[2]{V_0}{v}\right) \left(\pdv[2]{V_0}{x} \right) \right]^{-1} \left[ \left( \pdv[2]{V_0}{v} \right)\left( \pdv{V_1}{x} \right) - \left( \pdv{V_0}{v}{x} \right)\left( \pdv{V_1}{v} \right) \right].
\end{align}
All derivatives are to be evaluated at the tree-level minimum $(v_0, x_0)$.  Note that corrections to the VEVs contribute only at $\mathcal{O}(\hbar^2)$. This form of $\Veff(v_\text{min}, x_\text{min})$ is gauge invariant, and we shall calculate it in Landau gauge $\xi = 0$. With this choice, ghost fields remain massless after symmetry breaking and decouple from Goldstone modes.

From Eq.~(\ref{eq:Veff-definition}) we obtain $V_0(v,x) = V_\text{tree}(v,x)$, while the $\mathcal{O}(\hbar)$ part can be calculated by diagonalizing the quadratic Lagrangian in momentum space. In the $\mathcal{V} \rightarrow \infty$ limit, the result is the familiar Coleman-Weinberg correction in $d = 3-2\epsilon$ Euclidean dimensions:
\begin{align}
\label{eq:1loop-Veff}
V_1(v,x) = 2(d-1) J\left(m^2_W\right) + (d-1)J\left(m^2_Z\right) + 3J\left(m^2_1\right) + 2J\left(m^2_2\right) + J\left(m^2_+\right) + J\left(m^2_-\right).
\end{align}
Here the integral 
\begin{align}
J(m^2) = \frac12 \int \frac{d^d p}{(2\pi)^d} \ln \left( p^2 + m^2\right) = -\frac{(m^2)^{3/2}}{12\pi} + \mathcal{O}(\epsilon)
\end{align}
is finite in 3d, and the field-dependent masses read
\begin{align}
\label{eq:masses}
m^2_W =& \frac14 g^2 v^2 + g^2 x^2, \quad m^2_Z = \frac14(g^2 + g'^2)v^2, \nonumber \\
m^2_1 =& \mu^2_\phi + \lambda v^2 + \frac12 a_2 x^2, \quad m^2_2 = \mu^2_\Sigma + b_4 x^2 + \frac12 a_2 v^2, \nonumber \\
m^2_{\pm} =& \frac12 \left( m^2_3 + m^2_4 \pm \sqrt{(m^2_3 - m^2_4)^2 + 4 a_2^2 v^2 x^2} \right)
\end{align}
with 
\begin{align}
m^2_3 = \mu^2_\phi + 3\lambda v^2 + \frac12 a_2 x^2, \quad m^2_4 = \mu^2_\Sigma + 3 b_4 x^2 + \frac12 a_2 v^2.
\end{align}

The $\mathcal{O}(\hbar^2)$ correction consists of the 2-loop potential evaluated at a tree-level minimum, as well as 1-loop corrections to locations of the minima. The latter is obtained from Eqs.~(\ref{eq:min-corrections})-(\ref{eq:1loop-Veff}), while the former requires computation of one-particle-irreducible vacuum diagrams depicted in Fig.~\ref{fig:two-loop-diagrams}. Including the minus sign from $\exp\left[ -\frac{1}{\hbar} \int d^3x \; \mathcal{L}^{(I)}_\text{3d} \right]$ in the diagrammatic vertex rules, $V_2$ is given by minus the sum of diagrams in Fig.~\ref{fig:two-loop-diagrams}.

\begin{figure}
	\begin{center}
		\includegraphics[width=0.5\textwidth]{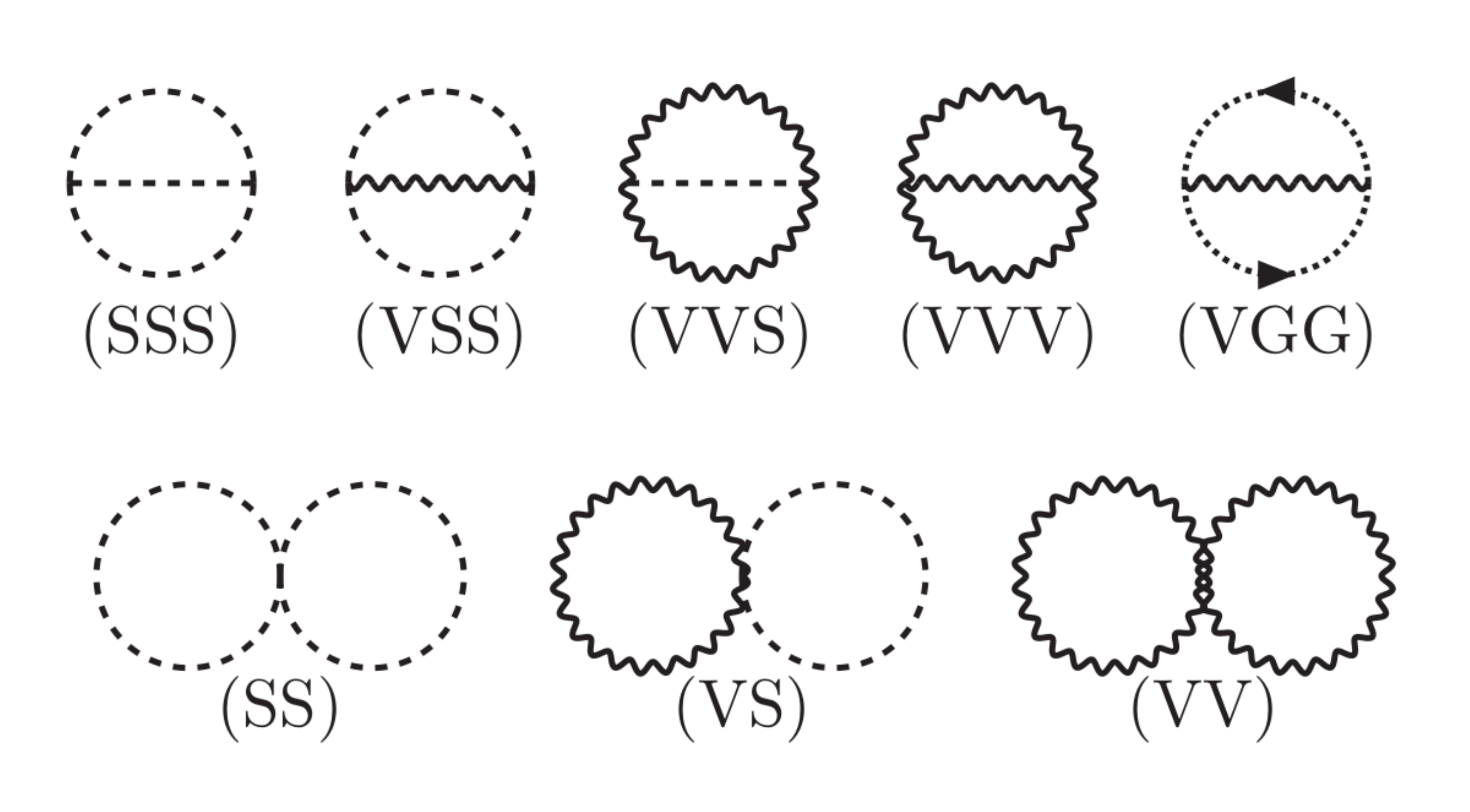}
	\end{center}
	\caption{Diagram topologies that enter the calculation of two-loop effective potential. Dashed lines denote scalars (S), wavy lines denote vector bosons (V) and dotted lines refer to ghost fields (G).}
	\label{fig:two-loop-diagrams}
\end{figure}

The calculation of $V_2(v,x)$ at general field values is somewhat complicated as one needs to introduce a field-dependent mixing angle for the neutral scalars. A simpler approach is to perform the computation directly at a tree-level minimum $(v_0, x_0)$, which is all that is needed for the $\mathcal{O}(\hbar^2)$ correction. In the case of $\Sigma$SM, there is then no mixing between the mass eigenstates of $\phi$ and $\Sigma$ as guaranteed by the $Z_2$ symmetry. Consequently, the masses $m^2_\pm$ in Eq.~(\ref{eq:masses}) reduce to $m_3^2$ and $m_4^2$. Below we present results for the different diagram topologies at two loops, expressed in terms of master integrals, but emphasize that these results are not applicable if $v_0$ and $x_0$ are simultaneously non-vanishing. Contributions from the $\Sigma$ field are collected in curly brackets.

\begin{align}
\label{eq:2loop-diags1}
\text{(SSS)} &= 3 \lambda^2 v^2_0 \mathcal{D}_{SSS}(m_3,m_1,m_1) + 3 \lambda^2 v^2_0 \mathcal{D}_{SSS}(m_3,m_3,m_3) \nonumber \\
& + \bigg\{ \frac{3}{4} a^2_2 x^2_0 \mathcal{D}_{SSS}(m_1,m_1,m_4) + \frac{1}{2} a^2_2 v^2_0 \mathcal{D}_{SSS}(m_2,m_2,m_3) + \frac{1}{4} a^2_2 x^2_0 \mathcal{D}_{SSS}(m_3,m_3,m_4) \nonumber \\
& + \frac{1}{4} a^2_2 v^2_0 \mathcal{D}_{SSS}(m_4,m_4,m_3) + 2 b^2_4 x^2_0 \mathcal{D}_{SSS}(m_2,m_2,m_4) + 3 b^2_4 x^2_0 \mathcal{D}_{SSS}(m_4,m_4,m_4) \bigg\}_{\Sigma SM} , \\ \nonumber \\
\text{(VSS)} &= \frac{1}{4} g^2  \mathcal{D}_{VSS}(m_1,m_1,m_W) + \frac{1}{4} g^2  \mathcal{D}_{VSS}(m_3,m_1,m_W) + \frac{1}{8} (g^2 + {g'}^2)  \mathcal{D}_{VSS}(m_3,m_1,m_Z) \nonumber \\
& + \frac{1}{8} \frac{(g^2 - {g'}^2)^2}{g^2 + {g'}^2} \mathcal{D}_{VSS}(m_1,m_1,m_Z) + \frac{1}{2} \frac{g^2 {g'}^2}{g^2 + {g'}^2} \mathcal{D}_{VSS}(m_1,m_1,0) \nonumber \\
& + \bigg\{ g^2 \mathcal{D}_{VSS}(m_4,m_2,m_W) + \frac{1}{2} \frac{g^4}{g^2 + {g'}^2} \mathcal{D}_{VSS}(m_2,m_2,m_Z) \nonumber \\
& + \frac{1}{2} \frac{g^2 {g'}^2}{g^2 + {g'}^2} \mathcal{D}_{VSS}(m_2,m_2,0)  \bigg\}_{\Sigma SM}, \\ \nonumber \\
\text{(VVS)} &= \frac{1}{8} g^4 v^2_0  \mathcal{D}_{VVS}(m_3,m_W,m_W) + \frac{1}{16} (g^2 + {g'}^2)^2 v^2_0  \mathcal{D}_{VVS}(m_3,m_Z,m_Z) \nonumber \\ 
& + \frac{1}{4} \frac{g^4 {g'}^2 v^2_0}{g^2 + {g'}^2} \mathcal{D}_{VVS}(m_1,m_W,0) + \frac{1}{4} \frac{g^2 {g'}^4 v^2_0}{g^2 + {g'}^2} \mathcal{D}_{VVS}(m_1,m_W,m_Z) \nonumber \\
& + \bigg\{ 2 g^4 x^2_0  \mathcal{D}_{VVS}(m_4,m_W,m_W) + \frac{g^6 x^2_0}{g^2 + {g'}^2} \mathcal{D}_{VVS}(m_2,m_W,m_Z) \nonumber \\
& +  \frac{g^4 {g'}^2 x^2_0}{g^2 + {g'}^2} \mathcal{D}_{VVS}(m_2,m_W,0)  \bigg\}_{\Sigma SM} , \\ \nonumber \\
\text{(VVV)} &= \frac{1}{2} \frac{g^4}{g^2+{g'}^2} \mathcal{D}_{VVV}(m_W,m_W,m_Z) +\frac{1}{2} \frac{g^2 {g'}^2}{g^2+{g'}^2} \mathcal{D}_{VVV}(m_W,m_W,0), \\ \nonumber \\
\text{(VGG)} &= -2 g^2 \mathcal{D}_{VGG}(m_W) -\frac{g^4}{g^2+{g'}^2} \mathcal{D}_{VGG}(m_Z) , \\ \nonumber \\
\text{(SS)} &= -\frac{15}{4} \lambda \Big( I^3_1(m_1) \Big)^2 - \frac{3}{2} \lambda I_1(m_1) I_1(m_3)   -\frac{3}{4} \lambda \Big( I_1(m_3) \Big)^2 \nonumber \\
& + \bigg\{ -\frac{3}{2} a_2  I_1(m_1) I_1(m_2)  -2 b_4 \Big( I_1(m_2) \Big)^2 -\frac{1}{2} a_2  I_1(m_2) I_1(m_3) -\frac{3}{4} a_2  I_1(m_1) I_1(m_4)  \nonumber \\
& - b_4  I_1(m_2) I_1(m_4) - \frac{1}{4} a_2  I_1(m_3) I_1(m_4) - \frac{3}{4} b_4 \Big( I_1(m_4) \Big)^2 \bigg\}_{\Sigma SM}, \\ \nonumber \\
\text{(VS)} &= - \frac{3}{4} (d-1) g^2  I_1(m_1) I_1(m_W) - \frac{1}{4} (d-1) \frac{(g^2-{g'}^2)^2}{g^2+{g'}^2}  I_1(m_1) I_1(m_Z) \nonumber \\
& - \frac{1}{4} (d-1) g^2  I_1(m_3) I_1(m_W) - \frac{1}{8} (d-1) (g^2 + {g'}^2)  I_1(m_1) I_1(m_Z) \nonumber \\
& - \frac{1}{8} (d-1) (g^2 + {g'}^2)  I_1(m_3) I_1(m_Z) \nonumber \\
&  + \bigg\{ -(d-1) g^2  I_1(m_2) I_1(m_W)  -(d-1) g^2  I_1(m_4) I_1(m_W) \nonumber \\
& - (d-1) \frac{g^4}{g^2+{g'}^2}  I_1(m_2) I_1(m_Z)  \bigg\}_{\Sigma SM}, \\ \nonumber \\
\label{eq:2loop-diags2}
\text{(VV)} &= -\frac{1}{2} g^2 \mathcal{D}_{VV}(m_W, m_W) -\frac{g^4}{g^2+{g'}^2}\mathcal{D}_{VV}(m_W, m_Z).
\end{align}
The loop integrals are defined, in dimensional regularization with \MSbar scale $\Lambda$, as 
\begin{align}
\int_p &\equiv \Big( \frac{e^{\gamma}\Lambda^2}{4\pi} \Big)^\epsilon \int \frac{d^d p}{(2\pi)^d}, \\ \nonumber \\
I_\alpha(m) &\equiv \int_p \frac{1}{(p^2+m^2)^\alpha} = \Big( \frac{e^{\gamma}\Lambda^2}{4\pi} \Big)^\epsilon \frac{(m^2)^{\frac{d}{2}-\alpha}}{(4\pi)^{\frac{d}{2}}} \frac{\Gamma(\alpha-\frac{d}{2})}{\Gamma(\alpha)}, \\ \nonumber \\
\mathcal{D}_{VV}(m_1,m_2) &\equiv \int_{p,k} \frac{ \delta_{ir} \delta_{js} + \delta_{ij} \delta_{rs} - 2 \delta_{is} \delta_{jr}}{(p^2+m^2_1)(k^2+m^2_2)} \Big( \delta_{ij} - \frac{p_i p_j}{p^2} \Big) \Big( \delta_{rs} - \frac{k_r k_s}{k^2} \Big) \nonumber \\
& = \frac{(d-1)^3}{d} I_1(m_1) I_1(m_2), \\ \nonumber \\
\mathcal{D}_{SSS}(m_1,m_2,m_3) &\equiv  \int_{p,k} \frac{1}{(p^2+m^2_1) (k^2+m^2_2) ((p+k)^2 + m^2_3) } \nonumber \\
& = \frac{1}{16\pi^2} \bigg(\frac{1}{4\epsilon} + \frac{1}{2} + \ln\Big( \frac{\Lambda}{m_1 + m_2 + m_3} \Big) \bigg) + \mathcal{O}(\epsilon) , \\ \nonumber \\
\mathcal{D}_{VGG}(m) &\equiv \int_{p,k} \frac{k_i (p+k)_j}{(p^2+m^2)(p+k)^2 k^2} \Big( \delta_{ij} - \frac{p_i p_j}{p^2} \Big) = \frac{1}{4}m^2 \mathcal{D}_{SSS}(m,0,0), \\ \nonumber \\
\mathcal{D}_{VSS}(m_1,m_2,m_3) &\equiv \int_{p,k} \frac{(2p+k)_i (2p+k)_j}{(p^2+m^2_1) (k^2+m^2_3) ((p+k)^2 + m^2_2)} \Big( \delta_{ij} - \frac{k_i k_j}{k^2} \Big), \\ \nonumber \\
\mathcal{D}_{VVS}(m_1,m_2,m_3) &\equiv \int_{p,k} \frac{ \delta_{ik} \delta_{jl} }{(p^2+m^2_2) (k^2+m^2_3) ((p+k)^2 + m^2_1)} \Big( \delta_{ij} - \frac{p_i p_j}{p^2} \Big) \Big( \delta_{kl} - \frac{k_k k_l}{k^2} \Big)  , \\ \nonumber \\
\mathcal{D}_{VVV}(m_1,m_2,m_3) &\equiv \int_{p,k} \frac{1}{(p^2+m^2_1) (k^2+m^2_2) ((p+k)^2 + m^2_3)} \Big( \delta_{ij} - \frac{p_i p_j}{p^2} \Big) \Big( \delta_{kl} - \frac{k_k k_l}{k^2} \Big) \Big( \delta_{rs} - \frac{(p+k)_r (p+k)_s}{(p+k)^2} \Big) \nonumber \\
& \times \Big( (2k + p)_i \delta_{rk} - (2k + p)_r \delta_{ik} + (k - p)_k \delta_{ir} \Big) 
         \Big( (k - p)_l \delta_{sj} - (2k + p)_s \delta_{lj} + (2p + k)_j \delta_{ls} \Big)
.
\end{align}
Some special cases of the vector ``sunset'' integrals have been calculated previously in \cite{Farakos:1994kx}. In the presence of the hypercharge gauge field, the following generalizations are needed. 
\begin{align}
\mathcal{D}_{VSS}(m_1,m_2,m_3) &= \frac{1}{m^2_3} \bigg( (-m^2_1 + m^2_2 + m^2_3) I_1(m_2)I_1(m_3) \nonumber \\
& + \Big( -m^2_3 I_1(m_2) + (m^2_1 - m^2_2 + m^2_3) I_1(m_3) \Big) I_1(m_1) -  (m^2_1 - m^2_2)^2 \mathcal{D}_{SSS}(m_1,m_2,0) \nonumber \\
& + (m_1 - m_2 - m_3)(m_1 + m_2 - m_3)(m_1 - m_2 + m_3)(m_1 + m_2 + m_3) \mathcal{D}_{SSS}(m_1,m_2,m_3)  \bigg) , \\ \nonumber \\
\mathcal{D}_{VSS}(m_1,m_2,0) & = -(d-1) \Big( (m^2_1 + m^2_2) \mathcal{D}_{SSS}(m_1,m_2,0) +  I_1(m_1) I_1(m_2) \Big) , \\ \nonumber \\
%%%%
%%%%
\mathcal{D}_{VVS}(m_1,m_2,m_3) &= \frac{1}{4 m^2_2 m^2_3} \bigg( - m^2_3 I_1(m_1) I_1(m_2) + \Big( -m^2_2 I_1(m_1) \nonumber \\
& + (-m^2_1 + m^2_2 + m^2_3) I_1(m_2)  \Big) I_1(m_3) + m^4_1 \mathcal{D}_{SSS}(m_1,0,0) \nonumber \\ 
& - (m^2_2-m^2_1)^2 \mathcal{D}_{SSS}(m_1,m_2,0)- (m^2_3-m^2_1)^2 \mathcal{D}_{SSS}(m_1,m_3,0) \nonumber \\
& + \Big( (m^2_2 - m^2_1)^2 + [-2m^2_1 + (4d-6)m^2_2] m^2_3 + m^4_3 \Big) \mathcal{D}_{SSS}(m_1,m_2,m_3)  \bigg) , \\ \nonumber \\
\mathcal{D}_{VVS}(m_1,m_2,0) &= -\frac{d-1}{4 m^2_2} \Big( (m^2_1 -3 m^2_2) \mathcal{D}_{SSS}(m_1,m_2,0) - m^2_1 \mathcal{D}_{SSS}(m_1,0,0) +  I_1(m_1) I_1(m_2) \Big)  , \\ \nonumber \\
\mathcal{D}_{VVS}(m,0,0) &= \frac{d(d-1)}{4} \mathcal{D}_{SSS}(m,0,0) , \\ \nonumber \\
\mathcal{D}_{VVS}(0,m,0) &= \frac{3d(d-1)}{4} \mathcal{D}_{SSS}(m,0,0) , \\ \nonumber \\
%%%%
%%%%
\mathcal{D}_{VVV}(m_1,m_1,m_2) &=  -\frac{d m^4_1 - (5d-4) m^2_1 m^2_2 - d(4d-7)m^4_2}{2 d m^2_1 m^2_2} I_1(m_1) I_1(m_2) \nonumber \\
& +  \frac{4(3d^2 - 4d -1)m^4_1 - 2d(4d-7) m^2_1 m^2_2 - d m^4_2}{4d m^4_1} \Big( I_1(m_1) \Big)^2 \nonumber \\
& - \frac{(m^2_1 - m^2_2)^2 \Big( m^4_1 + 2(2d-3) m^2_1 m^2_2 + m^4_2 \Big) }{2 m^4_1 m^2_2} \mathcal{D}_{SSS}(m_1,m_2,0) \nonumber \\
& - \frac{(4 m^2_1 - m^2_2) \Big( 4(d-1) m^4_1 + 4(2d-3) m^2_1 m^2_2 + m^4_2 \Big) }{4 m^4_1} \mathcal{D}_{SSS}(m_1,m_1,m_2) \nonumber \\
& + \frac{m^6_2}{4 m^4_1} \mathcal{D}_{SSS}(m_2,0,0) + \frac{m^4_1}{2 m^2_2} \mathcal{D}_{SSS}(m_1,0,0), \\ \nonumber \\
\mathcal{D}_{VVV}(m,m,0) &= \frac{5d^3 - 19 d^2 + 15 d +3}{(d-3)d} \Big( I_1(m) \Big)^2 - \frac{(3d-5)}{2} m^2 \mathcal{D}_{SSS}(m,0,0).
\end{align}
Many of the expressions above utilize integration-by-part techniques developed in \cite{Chetyrkin:1981qh,Laporta:2001dd} (for thermal sum-integrals, see \cite{Nishimura:2012ee}). 
%For an overview of these techniques, see \cite{Philipps thesis}. 
We are grateful to Philipp Schicho for providing particularly simple expressions for the special cases of $D_{VSS}, D_{VVS}$ and $D_{VVV}$.

The two-loop diagrams are UV divergent, but are regulated (apart from the vacuum divergence) by mass counterterms in the tree-level part (\ref{eq:3d-Veff-tree}). These are given by 
\begin{align}
\delta \mu^2_\phi =& -\frac{1}{16\pi^2}\frac{1}{4\epsilon} \bigg( \frac{39}{16}g^4 -\frac{5}{16}{g'}^4 - \frac{9}{8}g^2 {g'}^2 + 3 \lambda (3g^2 + {g'}^2)- 12\lambda^2 - \frac32 a^2_2 + 6 a_2 g^2  \bigg) \\ 
\delta \mu^2_{\Sigma} =& -\frac{1}{16\pi^2}\frac{1}{4\epsilon} \bigg(-g^4 + a_2 (3g^2 + {g'}^2) + 20 b_{4}g^2 - 2 a^2_{2} - 10 b^2_{4} \bigg),
\end{align}
which were also obtained independently in Ref.~\cite{Niemi:2018asa}. Due to super-renormalizability, there are no further corrections to the counterterms at higher loop orders. Apart from contributions from the triplet and the hypercharge field, the two-loop expressions in Eqs.~(\ref{eq:2loop-diags1})-(\ref{eq:2loop-diags2}) agree with those given in \cite{Farakos:1994kx} for an $\gr{SU(2)}$-Higgs theory.

To study the phase structure, we evaluate $\Veff(v_0, x_0)$ separately in the three phases,
\begin{align}
\Veff^\text{symm}(T) &= \Veff(0, 0), \quad\quad
\Veff^\phi(T) = \Veff(\sqrt{-\mu^2_\phi / \lambda}, 0), \quad\quad
\Veff^\Sigma(T) = \Veff(0, \sqrt{-\mu^2_\Sigma / b_4}),
\end{align}
and varying the temperature (which is now encapsuled in the 3d parameters). Not all of the above minima exist simultaneously at a given temperature; this needs to be checked separately. The condition for $T_c$ is that the value of $\Veff$ in any two minima is degenerate, \textit{e.g.} $\Veff^\Sigma(T_c) = \Veff^\phi(T_c)$ for $\Sigma\rightarrow \phi$ transitions. Latent heat is calculated from the 3d potential (which has units GeV$^3$) as 
\begin{align}
L = -T^2 \left[ \left(\pdv{\Veff}{T}\right)_\text{high-T phase} - \left(\pdv{\Veff}{T}\right)_\text{low-T phase} \right].
\end{align}
Finally, the scalar condensates are given by \cite{Farakos:1994xh}
\begin{align}
\langle \he\phi\phi \rangle = \pdv{\Veff}{\mu^2_\phi}, \quad \quad \frac12 \langle \Sigma^a \Sigma^a \rangle = \pdv{\Veff}{\mu^2_\Sigma}.
\end{align}

As discussed in the main text and in Refs.~\cite{Laine:1994zq,Farakos:1994xh}, the two-loop potential constructed here is not useful for studying 
thermodynamic properties near the critical temperature for transitions out of the 
symmetric phase $(v_0, x_0) = (0,0)$. The issue lies in Eq.~(\ref{eq:exp-Veff-min}), which assumes that the true minimum is related to the tree-level one through small perturbations. This assumption breaks down at temperatures close to $O \rightarrow \phi$ or $O \rightarrow \Sigma$ transitions, for which the tree-level condition for $T_c$ is that the thermally-corrected mass parameter vanishes, $\mu^2_\phi (T_c) = 0$ or $\mu^2_\Sigma (T_c) = 0$. Given that the high-$T$ expansion parameter is $\sim g^2 T \times \text{(mass scale)}^{-1}$, perturbation theory is unreliable
near $T_c$. In particular, there is an explicit divergence at two-loop order due to the vanishing scalar mass \cite{Laine:1994zq}. This problem does not arise for transitions between two broken phases ($\Sigma \rightarrow \phi$) as the tree-level masses need not vanish for such transitions.

One may hope to regulate the problem by giving up on the expansion of $(v_\text{min}, x_\text{min})$ altogether and solve for the minimum of $\Veff$ ``exactly'', as is frequently done in the literature. This automatically incorporates higher-order corrections of the VEVs into the potential. The downside is that these corrections also include uncancelled gauge dependence, and estimating the effects of this residual gauge dependence on final results is not a well-defined endeavor. 
Even if the resulting potential is free of spurious IR divergences, there is still no guarantee that the perturbative description near $T_c$ is reliable.

\end{widetext}

\bibliographystyle{apsrev4-1}

\bibliography{triplet2}

\begin{thebibliography}{64}%
\makeatletter
\providecommand \@ifxundefined [1]{%
 \@ifx{#1\undefined}
}%
\providecommand \@ifnum [1]{%
 \ifnum #1\expandafter \@firstoftwo
 \else \expandafter \@secondoftwo
 \fi
}%
\providecommand \@ifx [1]{%
 \ifx #1\expandafter \@firstoftwo
 \else \expandafter \@secondoftwo
 \fi
}%
\providecommand \natexlab [1]{#1}%
\providecommand \enquote  [1]{``#1''}%
\providecommand \bibnamefont  [1]{#1}%
\providecommand \bibfnamefont [1]{#1}%
\providecommand \citenamefont [1]{#1}%
\providecommand \href@noop [0]{\@secondoftwo}%
\providecommand \href [0]{\begingroup \@sanitize@url \@href}%
\providecommand \@href[1]{\@@startlink{#1}\@@href}%
\providecommand \@@href[1]{\endgroup#1\@@endlink}%
\providecommand \@sanitize@url [0]{\catcode `\\12\catcode `\$12\catcode
  `\&12\catcode `\#12\catcode `\^12\catcode `\_12\catcode `\%12\relax}%
\providecommand \@@startlink[1]{}%
\providecommand \@@endlink[0]{}%
\providecommand \url  [0]{\begingroup\@sanitize@url \@url }%
\providecommand \@url [1]{\endgroup\@href {#1}{\urlprefix }}%
\providecommand \urlprefix  [0]{URL }%
\providecommand \Eprint [0]{\href }%
\providecommand \doibase [0]{http://dx.doi.org/}%
\providecommand \selectlanguage [0]{\@gobble}%
\providecommand \bibinfo  [0]{\@secondoftwo}%
\providecommand \bibfield  [0]{\@secondoftwo}%
\providecommand \translation [1]{[#1]}%
\providecommand \BibitemOpen [0]{}%
\providecommand \bibitemStop [0]{}%
\providecommand \bibitemNoStop [0]{.\EOS\space}%
\providecommand \EOS [0]{\spacefactor3000\relax}%
\providecommand \BibitemShut  [1]{\csname bibitem#1\endcsname}%
\let\auto@bib@innerbib\@empty
%</preamble>
\bibitem [{\citenamefont {Kajantie}\ \emph
  {et~al.}(1996{\natexlab{a}})\citenamefont {Kajantie}, \citenamefont {Laine},
  \citenamefont {Rummukainen},\ and\ \citenamefont
  {Shaposhnikov}}]{Kajantie:1996mn}%
  \BibitemOpen
  \bibfield  {author} {\bibinfo {author} {\bibfnamefont {K.}~\bibnamefont
  {Kajantie}}, \bibinfo {author} {\bibfnamefont {M.}~\bibnamefont {Laine}},
  \bibinfo {author} {\bibfnamefont {K.}~\bibnamefont {Rummukainen}}, \ and\
  \bibinfo {author} {\bibfnamefont {M.~E.}\ \bibnamefont {Shaposhnikov}},\
  }\href {\doibase 10.1103/PhysRevLett.77.2887} {\bibfield  {journal} {\bibinfo
   {journal} {Phys. Rev. Lett.}\ }\textbf {\bibinfo {volume} {77}},\ \bibinfo
  {pages} {2887} (\bibinfo {year} {1996}{\natexlab{a}})},\ \Eprint
  {http://arxiv.org/abs/hep-ph/9605288} {arXiv:hep-ph/9605288 [hep-ph]}
  \BibitemShut {NoStop}%
%%CITATION = HEP-PH/9605288;%%
\bibitem [{\citenamefont {Csikor}\ \emph {et~al.}(1999)\citenamefont {Csikor},
  \citenamefont {Fodor},\ and\ \citenamefont {Heitger}}]{Csikor:1998eu}%
  \BibitemOpen
  \bibfield  {author} {\bibinfo {author} {\bibfnamefont {F.}~\bibnamefont
  {Csikor}}, \bibinfo {author} {\bibfnamefont {Z.}~\bibnamefont {Fodor}}, \
  and\ \bibinfo {author} {\bibfnamefont {J.}~\bibnamefont {Heitger}},\ }\href
  {\doibase 10.1103/PhysRevLett.82.21} {\bibfield  {journal} {\bibinfo
  {journal} {Phys. Rev. Lett.}\ }\textbf {\bibinfo {volume} {82}},\ \bibinfo
  {pages} {21} (\bibinfo {year} {1999})},\ \Eprint
  {http://arxiv.org/abs/hep-ph/9809291} {arXiv:hep-ph/9809291 [hep-ph]}
  \BibitemShut {NoStop}%
%%CITATION = HEP-PH/9809291;%%
\bibitem [{\citenamefont {D'Onofrio}\ and\ \citenamefont
  {Rummukainen}(2016)}]{DOnofrio:2015gop}%
  \BibitemOpen
  \bibfield  {author} {\bibinfo {author} {\bibfnamefont {M.}~\bibnamefont
  {D'Onofrio}}\ and\ \bibinfo {author} {\bibfnamefont {K.}~\bibnamefont
  {Rummukainen}},\ }\href {\doibase 10.1103/PhysRevD.93.025003} {\bibfield
  {journal} {\bibinfo  {journal} {Phys. Rev.}\ }\textbf {\bibinfo {volume}
  {D93}},\ \bibinfo {pages} {025003} (\bibinfo {year} {2016})},\ \Eprint
  {http://arxiv.org/abs/1508.07161} {arXiv:1508.07161 [hep-ph]} \BibitemShut
  {NoStop}%
%%CITATION = ARXIV:1508.07161;%%
\bibitem [{\citenamefont {Kuzmin}\ \emph {et~al.}(1985)\citenamefont {Kuzmin},
  \citenamefont {Rubakov},\ and\ \citenamefont {Shaposhnikov}}]{Kuzmin:1985mm}%
  \BibitemOpen
  \bibfield  {author} {\bibinfo {author} {\bibfnamefont {V.~A.}\ \bibnamefont
  {Kuzmin}}, \bibinfo {author} {\bibfnamefont {V.~A.}\ \bibnamefont {Rubakov}},
  \ and\ \bibinfo {author} {\bibfnamefont {M.~E.}\ \bibnamefont
  {Shaposhnikov}},\ }\href {\doibase 10.1016/0370-2693(85)91028-7} {\bibfield
  {journal} {\bibinfo  {journal} {Phys. Lett.}\ }\textbf {\bibinfo {volume}
  {155B}},\ \bibinfo {pages} {36} (\bibinfo {year} {1985})}\BibitemShut
  {NoStop}%
%%CITATION = PHLTA,155B,36;%%
\bibitem [{\citenamefont {Shaposhnikov}(1986)}]{Shaposhnikov:1986jp}%
  \BibitemOpen
  \bibfield  {author} {\bibinfo {author} {\bibfnamefont {M.~E.}\ \bibnamefont
  {Shaposhnikov}},\ }\href@noop {} {\bibfield  {journal} {\bibinfo  {journal}
  {JETP Lett.}\ }\textbf {\bibinfo {volume} {44}},\ \bibinfo {pages} {465}
  (\bibinfo {year} {1986})},\ \bibinfo {note} {[Pisma Zh. Eksp. Teor.
  Fiz.44,364(1986)]}\BibitemShut {NoStop}%
%%CITATION = JTPLA,44,465;%%
\bibitem [{\citenamefont {Shaposhnikov}(1987)}]{Shaposhnikov:1987tw}%
  \BibitemOpen
  \bibfield  {author} {\bibinfo {author} {\bibfnamefont {M.~E.}\ \bibnamefont
  {Shaposhnikov}},\ }\href {\doibase 10.1016/0550-3213(87)90127-1} {\bibfield
  {journal} {\bibinfo  {journal} {Nucl. Phys.}\ }\textbf {\bibinfo {volume}
  {B287}},\ \bibinfo {pages} {757} (\bibinfo {year} {1987})}\BibitemShut
  {NoStop}%
%%CITATION = NUPHA,B287,757;%%
\bibitem [{\citenamefont {Ramsey-Musolf}(2020)}]{Ramsey-Musolf:2019lsf}%
  \BibitemOpen
  \bibfield  {author} {\bibinfo {author} {\bibfnamefont {M.~J.}\ \bibnamefont
  {Ramsey-Musolf}},\ }\href {\doibase 10.1007/JHEP09(2020)179} {\bibfield
  {journal} {\bibinfo  {journal} {JHEP}\ }\textbf {\bibinfo {volume} {09}},\
  \bibinfo {pages} {179} (\bibinfo {year} {2020})},\ \Eprint
  {http://arxiv.org/abs/1912.07189} {arXiv:1912.07189 [hep-ph]} \BibitemShut
  {NoStop}%
\bibitem [{\citenamefont {Caprini}\ \emph {et~al.}(2016)\citenamefont {Caprini}
  \emph {et~al.}}]{Caprini:2015zlo}%
  \BibitemOpen
  \bibfield  {author} {\bibinfo {author} {\bibfnamefont {C.}~\bibnamefont
  {Caprini}} \emph {et~al.},\ }\href {\doibase 10.1088/1475-7516/2016/04/001}
  {\bibfield  {journal} {\bibinfo  {journal} {JCAP}\ }\textbf {\bibinfo
  {volume} {1604}},\ \bibinfo {pages} {001} (\bibinfo {year} {2016})},\ \Eprint
  {http://arxiv.org/abs/1512.06239} {arXiv:1512.06239 [astro-ph.CO]}
  \BibitemShut {NoStop}%
%%CITATION = ARXIV:1512.06239;%%
\bibitem [{\citenamefont {Huang}\ \emph {et~al.}(2016)\citenamefont {Huang},
  \citenamefont {Wan}, \citenamefont {Wang}, \citenamefont {Cai},\ and\
  \citenamefont {Zhang}}]{Huang:2016odd}%
  \BibitemOpen
  \bibfield  {author} {\bibinfo {author} {\bibfnamefont {F.~P.}\ \bibnamefont
  {Huang}}, \bibinfo {author} {\bibfnamefont {Y.}~\bibnamefont {Wan}}, \bibinfo
  {author} {\bibfnamefont {D.-G.}\ \bibnamefont {Wang}}, \bibinfo {author}
  {\bibfnamefont {Y.-F.}\ \bibnamefont {Cai}}, \ and\ \bibinfo {author}
  {\bibfnamefont {X.}~\bibnamefont {Zhang}},\ }\href {\doibase
  10.1103/PhysRevD.94.041702} {\bibfield  {journal} {\bibinfo  {journal} {Phys.
  Rev. D}\ }\textbf {\bibinfo {volume} {94}},\ \bibinfo {pages} {041702}
  (\bibinfo {year} {2016})},\ \Eprint {http://arxiv.org/abs/1601.01640}
  {arXiv:1601.01640 [hep-ph]} \BibitemShut {NoStop}%
\bibitem [{\citenamefont {Caprini}\ \emph {et~al.}(2020)\citenamefont {Caprini}
  \emph {et~al.}}]{Caprini:2019egz}%
  \BibitemOpen
  \bibfield  {author} {\bibinfo {author} {\bibfnamefont {C.}~\bibnamefont
  {Caprini}} \emph {et~al.},\ }\href {\doibase 10.1088/1475-7516/2020/03/024}
  {\bibfield  {journal} {\bibinfo  {journal} {JCAP}\ }\textbf {\bibinfo
  {volume} {03}},\ \bibinfo {pages} {024} (\bibinfo {year} {2020})},\ \Eprint
  {http://arxiv.org/abs/1910.13125} {arXiv:1910.13125 [astro-ph.CO]}
  \BibitemShut {NoStop}%
\bibitem [{\citenamefont {Hammerschmitt}\ \emph {et~al.}(1994)\citenamefont
  {Hammerschmitt}, \citenamefont {Kripfganz},\ and\ \citenamefont
  {Schmidt}}]{Hammerschmitt:1994fn}%
  \BibitemOpen
  \bibfield  {author} {\bibinfo {author} {\bibfnamefont {A.}~\bibnamefont
  {Hammerschmitt}}, \bibinfo {author} {\bibfnamefont {J.}~\bibnamefont
  {Kripfganz}}, \ and\ \bibinfo {author} {\bibfnamefont {M.}~\bibnamefont
  {Schmidt}},\ }\href {\doibase 10.1007/BF01557241} {\bibfield  {journal}
  {\bibinfo  {journal} {Z. Phys. C}\ }\textbf {\bibinfo {volume} {64}},\
  \bibinfo {pages} {105} (\bibinfo {year} {1994})},\ \Eprint
  {http://arxiv.org/abs/hep-ph/9404272} {arXiv:hep-ph/9404272} \BibitemShut
  {NoStop}%
\bibitem [{\citenamefont {Patel}\ and\ \citenamefont
  {Ramsey-Musolf}(2013)}]{Patel:2012pi}%
  \BibitemOpen
  \bibfield  {author} {\bibinfo {author} {\bibfnamefont {H.~H.}\ \bibnamefont
  {Patel}}\ and\ \bibinfo {author} {\bibfnamefont {M.~J.}\ \bibnamefont
  {Ramsey-Musolf}},\ }\href {\doibase 10.1103/PhysRevD.88.035013} {\bibfield
  {journal} {\bibinfo  {journal} {Phys. Rev.}\ }\textbf {\bibinfo {volume}
  {D88}},\ \bibinfo {pages} {035013} (\bibinfo {year} {2013})},\ \Eprint
  {http://arxiv.org/abs/1212.5652} {arXiv:1212.5652 [hep-ph]} \BibitemShut
  {NoStop}%
%%CITATION = ARXIV:1212.5652;%%
\bibitem [{\citenamefont {Inoue}\ \emph {et~al.}(2016)\citenamefont {Inoue},
  \citenamefont {Ovanesyan},\ and\ \citenamefont
  {Ramsey-Musolf}}]{Inoue:2015pza}%
  \BibitemOpen
  \bibfield  {author} {\bibinfo {author} {\bibfnamefont {S.}~\bibnamefont
  {Inoue}}, \bibinfo {author} {\bibfnamefont {G.}~\bibnamefont {Ovanesyan}}, \
  and\ \bibinfo {author} {\bibfnamefont {M.~J.}\ \bibnamefont
  {Ramsey-Musolf}},\ }\href {\doibase 10.1103/PhysRevD.93.015013} {\bibfield
  {journal} {\bibinfo  {journal} {Phys. Rev.}\ }\textbf {\bibinfo {volume}
  {D93}},\ \bibinfo {pages} {015013} (\bibinfo {year} {2016})},\ \Eprint
  {http://arxiv.org/abs/1508.05404} {arXiv:1508.05404 [hep-ph]} \BibitemShut
  {NoStop}%
%%CITATION = ARXIV:1508.05404;%%
\bibitem [{\citenamefont {Blinov}\ \emph {et~al.}(2015)\citenamefont {Blinov},
  \citenamefont {Kozaczuk}, \citenamefont {Morrissey},\ and\ \citenamefont
  {Tamarit}}]{Blinov:2015sna}%
  \BibitemOpen
  \bibfield  {author} {\bibinfo {author} {\bibfnamefont {N.}~\bibnamefont
  {Blinov}}, \bibinfo {author} {\bibfnamefont {J.}~\bibnamefont {Kozaczuk}},
  \bibinfo {author} {\bibfnamefont {D.~E.}\ \bibnamefont {Morrissey}}, \ and\
  \bibinfo {author} {\bibfnamefont {C.}~\bibnamefont {Tamarit}},\ }\href
  {\doibase 10.1103/PhysRevD.92.035012} {\bibfield  {journal} {\bibinfo
  {journal} {Phys. Rev.}\ }\textbf {\bibinfo {volume} {D92}},\ \bibinfo {pages}
  {035012} (\bibinfo {year} {2015})},\ \Eprint
  {http://arxiv.org/abs/1504.05195} {arXiv:1504.05195 [hep-ph]} \BibitemShut
  {NoStop}%
%%CITATION = ARXIV:1504.05195;%%
\bibitem [{\citenamefont {Ramsey-Musolf}\ \emph {et~al.}(2018)\citenamefont
  {Ramsey-Musolf}, \citenamefont {Winslow},\ and\ \citenamefont
  {White}}]{Ramsey-Musolf:2017tgh}%
  \BibitemOpen
  \bibfield  {author} {\bibinfo {author} {\bibfnamefont {M.~J.}\ \bibnamefont
  {Ramsey-Musolf}}, \bibinfo {author} {\bibfnamefont {P.}~\bibnamefont
  {Winslow}}, \ and\ \bibinfo {author} {\bibfnamefont {G.}~\bibnamefont
  {White}},\ }\href {\doibase 10.1103/PhysRevD.97.123509} {\bibfield  {journal}
  {\bibinfo  {journal} {Phys. Rev. D}\ }\textbf {\bibinfo {volume} {97}},\
  \bibinfo {pages} {123509} (\bibinfo {year} {2018})},\ \Eprint
  {http://arxiv.org/abs/1708.07511} {arXiv:1708.07511 [hep-ph]} \BibitemShut
  {NoStop}%
\bibitem [{\citenamefont {Jeannerot}\ \emph {et~al.}(2003)\citenamefont
  {Jeannerot}, \citenamefont {Rocher},\ and\ \citenamefont
  {Sakellariadou}}]{Jeannerot:2003qv}%
  \BibitemOpen
  \bibfield  {author} {\bibinfo {author} {\bibfnamefont {R.}~\bibnamefont
  {Jeannerot}}, \bibinfo {author} {\bibfnamefont {J.}~\bibnamefont {Rocher}}, \
  and\ \bibinfo {author} {\bibfnamefont {M.}~\bibnamefont {Sakellariadou}},\
  }\href {\doibase 10.1103/PhysRevD.68.103514} {\bibfield  {journal} {\bibinfo
  {journal} {Phys. Rev. D}\ }\textbf {\bibinfo {volume} {68}},\ \bibinfo
  {pages} {103514} (\bibinfo {year} {2003})},\ \Eprint
  {http://arxiv.org/abs/hep-ph/0308134} {arXiv:hep-ph/0308134} \BibitemShut
  {NoStop}%
\bibitem [{\citenamefont {Zurek}(1996)}]{Zurek:1996sj}%
  \BibitemOpen
  \bibfield  {author} {\bibinfo {author} {\bibfnamefont {W.}~\bibnamefont
  {Zurek}},\ }\href {\doibase 10.1016/S0370-1573(96)00009-9} {\bibfield
  {journal} {\bibinfo  {journal} {Phys. Rept.}\ }\textbf {\bibinfo {volume}
  {276}},\ \bibinfo {pages} {177} (\bibinfo {year} {1996})},\ \Eprint
  {http://arxiv.org/abs/cond-mat/9607135} {arXiv:cond-mat/9607135} \BibitemShut
  {NoStop}%
\bibitem [{\citenamefont {Profumo}\ \emph {et~al.}(2007)\citenamefont
  {Profumo}, \citenamefont {Ramsey-Musolf},\ and\ \citenamefont
  {Shaughnessy}}]{Profumo:2007wc}%
  \BibitemOpen
  \bibfield  {author} {\bibinfo {author} {\bibfnamefont {S.}~\bibnamefont
  {Profumo}}, \bibinfo {author} {\bibfnamefont {M.~J.}\ \bibnamefont
  {Ramsey-Musolf}}, \ and\ \bibinfo {author} {\bibfnamefont {G.}~\bibnamefont
  {Shaughnessy}},\ }\href {\doibase 10.1088/1126-6708/2007/08/010} {\bibfield
  {journal} {\bibinfo  {journal} {JHEP}\ }\textbf {\bibinfo {volume} {08}},\
  \bibinfo {pages} {010} (\bibinfo {year} {2007})},\ \Eprint
  {http://arxiv.org/abs/0705.2425} {arXiv:0705.2425 [hep-ph]} \BibitemShut
  {NoStop}%
\bibitem [{\citenamefont {Espinosa}\ \emph {et~al.}(2012)\citenamefont
  {Espinosa}, \citenamefont {Konstandin},\ and\ \citenamefont
  {Riva}}]{Espinosa:2011ax}%
  \BibitemOpen
  \bibfield  {author} {\bibinfo {author} {\bibfnamefont {J.~R.}\ \bibnamefont
  {Espinosa}}, \bibinfo {author} {\bibfnamefont {T.}~\bibnamefont
  {Konstandin}}, \ and\ \bibinfo {author} {\bibfnamefont {F.}~\bibnamefont
  {Riva}},\ }\href {\doibase 10.1016/j.nuclphysb.2011.09.010} {\bibfield
  {journal} {\bibinfo  {journal} {Nucl. Phys. B}\ }\textbf {\bibinfo {volume}
  {854}},\ \bibinfo {pages} {592} (\bibinfo {year} {2012})},\ \Eprint
  {http://arxiv.org/abs/1107.5441} {arXiv:1107.5441 [hep-ph]} \BibitemShut
  {NoStop}%
\bibitem [{\citenamefont {Patel}\ \emph {et~al.}(2013)\citenamefont {Patel},
  \citenamefont {Ramsey-Musolf},\ and\ \citenamefont {Wise}}]{Patel:2013zla}%
  \BibitemOpen
  \bibfield  {author} {\bibinfo {author} {\bibfnamefont {H.~H.}\ \bibnamefont
  {Patel}}, \bibinfo {author} {\bibfnamefont {M.~J.}\ \bibnamefont
  {Ramsey-Musolf}}, \ and\ \bibinfo {author} {\bibfnamefont {M.~B.}\
  \bibnamefont {Wise}},\ }\href {\doibase 10.1103/PhysRevD.88.015003}
  {\bibfield  {journal} {\bibinfo  {journal} {Phys. Rev. D}\ }\textbf {\bibinfo
  {volume} {88}},\ \bibinfo {pages} {015003} (\bibinfo {year} {2013})},\
  \Eprint {http://arxiv.org/abs/1303.1140} {arXiv:1303.1140 [hep-ph]}
  \BibitemShut {NoStop}%
\bibitem [{\citenamefont {Curtin}\ \emph {et~al.}(2014)\citenamefont {Curtin},
  \citenamefont {Meade},\ and\ \citenamefont {Yu}}]{Curtin:2014jma}%
  \BibitemOpen
  \bibfield  {author} {\bibinfo {author} {\bibfnamefont {D.}~\bibnamefont
  {Curtin}}, \bibinfo {author} {\bibfnamefont {P.}~\bibnamefont {Meade}}, \
  and\ \bibinfo {author} {\bibfnamefont {C.-T.}\ \bibnamefont {Yu}},\ }\href
  {\doibase 10.1007/JHEP11(2014)127} {\bibfield  {journal} {\bibinfo  {journal}
  {JHEP}\ }\textbf {\bibinfo {volume} {11}},\ \bibinfo {pages} {127} (\bibinfo
  {year} {2014})},\ \Eprint {http://arxiv.org/abs/1409.0005} {arXiv:1409.0005
  [hep-ph]} \BibitemShut {NoStop}%
\bibitem [{\citenamefont {Jiang}\ \emph {et~al.}(2016)\citenamefont {Jiang},
  \citenamefont {Bian}, \citenamefont {Huang},\ and\ \citenamefont
  {Shu}}]{Jiang:2015cwa}%
  \BibitemOpen
  \bibfield  {author} {\bibinfo {author} {\bibfnamefont {M.}~\bibnamefont
  {Jiang}}, \bibinfo {author} {\bibfnamefont {L.}~\bibnamefont {Bian}},
  \bibinfo {author} {\bibfnamefont {W.}~\bibnamefont {Huang}}, \ and\ \bibinfo
  {author} {\bibfnamefont {J.}~\bibnamefont {Shu}},\ }\href {\doibase
  10.1103/PhysRevD.93.065032} {\bibfield  {journal} {\bibinfo  {journal} {Phys.
  Rev. D}\ }\textbf {\bibinfo {volume} {93}},\ \bibinfo {pages} {065032}
  (\bibinfo {year} {2016})},\ \Eprint {http://arxiv.org/abs/1502.07574}
  {arXiv:1502.07574 [hep-ph]} \BibitemShut {NoStop}%
\bibitem [{\citenamefont {Kurup}\ and\ \citenamefont
  {Perelstein}(2017)}]{Kurup:2017dzf}%
  \BibitemOpen
  \bibfield  {author} {\bibinfo {author} {\bibfnamefont {G.}~\bibnamefont
  {Kurup}}\ and\ \bibinfo {author} {\bibfnamefont {M.}~\bibnamefont
  {Perelstein}},\ }\href {\doibase 10.1103/PhysRevD.96.015036} {\bibfield
  {journal} {\bibinfo  {journal} {Phys. Rev.}\ }\textbf {\bibinfo {volume}
  {D96}},\ \bibinfo {pages} {015036} (\bibinfo {year} {2017})},\ \Eprint
  {http://arxiv.org/abs/1704.03381} {arXiv:1704.03381 [hep-ph]} \BibitemShut
  {NoStop}%
%%CITATION = ARXIV:1704.03381;%%
\bibitem [{\citenamefont {Chiang}\ \emph {et~al.}(2018)\citenamefont {Chiang},
  \citenamefont {Ramsey-Musolf},\ and\ \citenamefont
  {Senaha}}]{Chiang:2017nmu}%
  \BibitemOpen
  \bibfield  {author} {\bibinfo {author} {\bibfnamefont {C.-W.}\ \bibnamefont
  {Chiang}}, \bibinfo {author} {\bibfnamefont {M.~J.}\ \bibnamefont
  {Ramsey-Musolf}}, \ and\ \bibinfo {author} {\bibfnamefont {E.}~\bibnamefont
  {Senaha}},\ }\href {\doibase 10.1103/PhysRevD.97.015005} {\bibfield
  {journal} {\bibinfo  {journal} {Phys. Rev. D}\ }\textbf {\bibinfo {volume}
  {97}},\ \bibinfo {pages} {015005} (\bibinfo {year} {2018})},\ \Eprint
  {http://arxiv.org/abs/1707.09960} {arXiv:1707.09960 [hep-ph]} \BibitemShut
  {NoStop}%
\bibitem [{\citenamefont {Kang}\ \emph {et~al.}(2018)\citenamefont {Kang},
  \citenamefont {Ko},\ and\ \citenamefont {Matsui}}]{Kang:2017mkl}%
  \BibitemOpen
  \bibfield  {author} {\bibinfo {author} {\bibfnamefont {Z.}~\bibnamefont
  {Kang}}, \bibinfo {author} {\bibfnamefont {P.}~\bibnamefont {Ko}}, \ and\
  \bibinfo {author} {\bibfnamefont {T.}~\bibnamefont {Matsui}},\ }\href
  {\doibase 10.1007/JHEP02(2018)115} {\bibfield  {journal} {\bibinfo  {journal}
  {JHEP}\ }\textbf {\bibinfo {volume} {02}},\ \bibinfo {pages} {115} (\bibinfo
  {year} {2018})},\ \Eprint {http://arxiv.org/abs/1706.09721} {arXiv:1706.09721
  [hep-ph]} \BibitemShut {NoStop}%
\bibitem [{\citenamefont {Linde}(1980)}]{Linde:1980ts}%
  \BibitemOpen
  \bibfield  {author} {\bibinfo {author} {\bibfnamefont {A.~D.}\ \bibnamefont
  {Linde}},\ }\href {\doibase 10.1016/0370-2693(80)90769-8} {\bibfield
  {journal} {\bibinfo  {journal} {Phys. Lett.}\ }\textbf {\bibinfo {volume}
  {96B}},\ \bibinfo {pages} {289} (\bibinfo {year} {1980})}\BibitemShut
  {NoStop}%
%%CITATION = PHLTA,96B,289;%%
\bibitem [{\citenamefont {Laine}\ \emph {et~al.}(2017)\citenamefont {Laine},
  \citenamefont {Meyer},\ and\ \citenamefont {Nardini}}]{Laine:2017hdk}%
  \BibitemOpen
  \bibfield  {author} {\bibinfo {author} {\bibfnamefont {M.}~\bibnamefont
  {Laine}}, \bibinfo {author} {\bibfnamefont {M.}~\bibnamefont {Meyer}}, \ and\
  \bibinfo {author} {\bibfnamefont {G.}~\bibnamefont {Nardini}},\ }\href
  {\doibase 10.1016/j.nuclphysb.2017.04.023} {\bibfield  {journal} {\bibinfo
  {journal} {Nucl. Phys.}\ }\textbf {\bibinfo {volume} {B920}},\ \bibinfo
  {pages} {565} (\bibinfo {year} {2017})},\ \Eprint
  {http://arxiv.org/abs/1702.07479} {arXiv:1702.07479 [hep-ph]} \BibitemShut
  {NoStop}%
%%CITATION = ARXIV:1702.07479;%%
\bibitem [{\citenamefont {Kainulainen}\ \emph {et~al.}(2019)\citenamefont
  {Kainulainen}, \citenamefont {Keus}, \citenamefont {Niemi}, \citenamefont
  {Rummukainen}, \citenamefont {Tenkanen},\ and\ \citenamefont
  {Vaskonen}}]{Kainulainen:2019kyp}%
  \BibitemOpen
  \bibfield  {author} {\bibinfo {author} {\bibfnamefont {K.}~\bibnamefont
  {Kainulainen}}, \bibinfo {author} {\bibfnamefont {V.}~\bibnamefont {Keus}},
  \bibinfo {author} {\bibfnamefont {L.}~\bibnamefont {Niemi}}, \bibinfo
  {author} {\bibfnamefont {K.}~\bibnamefont {Rummukainen}}, \bibinfo {author}
  {\bibfnamefont {T.~V.~I.}\ \bibnamefont {Tenkanen}}, \ and\ \bibinfo {author}
  {\bibfnamefont {V.}~\bibnamefont {Vaskonen}},\ }\href {\doibase
  10.1007/JHEP06(2019)075} {\bibfield  {journal} {\bibinfo  {journal} {JHEP}\
  }\textbf {\bibinfo {volume} {06}},\ \bibinfo {pages} {075} (\bibinfo {year}
  {2019})},\ \Eprint {http://arxiv.org/abs/1904.01329} {arXiv:1904.01329
  [hep-ph]} \BibitemShut {NoStop}%
%%CITATION = ARXIV:1904.01329;%%
\bibitem [{\citenamefont {Ginsparg}(1980)}]{Ginsparg:1980ef}%
  \BibitemOpen
  \bibfield  {author} {\bibinfo {author} {\bibfnamefont {P.~H.}\ \bibnamefont
  {Ginsparg}},\ }\href {\doibase 10.1016/0550-3213(80)90418-6} {\bibfield
  {journal} {\bibinfo  {journal} {Nucl. Phys.}\ }\textbf {\bibinfo {volume}
  {B170}},\ \bibinfo {pages} {388} (\bibinfo {year} {1980})}\BibitemShut
  {NoStop}%
%%CITATION = NUPHA,B170,388;%%
\bibitem [{\citenamefont {Appelquist}\ and\ \citenamefont
  {Pisarski}(1981)}]{Appelquist:1981vg}%
  \BibitemOpen
  \bibfield  {author} {\bibinfo {author} {\bibfnamefont {T.}~\bibnamefont
  {Appelquist}}\ and\ \bibinfo {author} {\bibfnamefont {R.~D.}\ \bibnamefont
  {Pisarski}},\ }\href {\doibase 10.1103/PhysRevD.23.2305} {\bibfield
  {journal} {\bibinfo  {journal} {Phys. Rev.}\ }\textbf {\bibinfo {volume}
  {D23}},\ \bibinfo {pages} {2305} (\bibinfo {year} {1981})}\BibitemShut
  {NoStop}%
%%CITATION = PHRVA,D23,2305;%%
\bibitem [{\citenamefont {Kajantie}\ \emph
  {et~al.}(1996{\natexlab{b}})\citenamefont {Kajantie}, \citenamefont {Laine},
  \citenamefont {Rummukainen},\ and\ \citenamefont
  {Shaposhnikov}}]{Kajantie:1995dw}%
  \BibitemOpen
  \bibfield  {author} {\bibinfo {author} {\bibfnamefont {K.}~\bibnamefont
  {Kajantie}}, \bibinfo {author} {\bibfnamefont {M.}~\bibnamefont {Laine}},
  \bibinfo {author} {\bibfnamefont {K.}~\bibnamefont {Rummukainen}}, \ and\
  \bibinfo {author} {\bibfnamefont {M.~E.}\ \bibnamefont {Shaposhnikov}},\
  }\href {\doibase 10.1016/0550-3213(95)00549-8} {\bibfield  {journal}
  {\bibinfo  {journal} {Nucl. Phys.}\ }\textbf {\bibinfo {volume} {B458}},\
  \bibinfo {pages} {90} (\bibinfo {year} {1996}{\natexlab{b}})},\ \Eprint
  {http://arxiv.org/abs/hep-ph/9508379} {arXiv:hep-ph/9508379 [hep-ph]}
  \BibitemShut {NoStop}%
%%CITATION = HEP-PH/9508379;%%
\bibitem [{\citenamefont {Fileviez~Perez}\ \emph {et~al.}(2009)\citenamefont
  {Fileviez~Perez}, \citenamefont {Patel}, \citenamefont {Ramsey-Musolf},\ and\
  \citenamefont {Wang}}]{FileviezPerez:2008bj}%
  \BibitemOpen
  \bibfield  {author} {\bibinfo {author} {\bibfnamefont {P.}~\bibnamefont
  {Fileviez~Perez}}, \bibinfo {author} {\bibfnamefont {H.~H.}\ \bibnamefont
  {Patel}}, \bibinfo {author} {\bibfnamefont {M.}~\bibnamefont
  {Ramsey-Musolf}}, \ and\ \bibinfo {author} {\bibfnamefont {K.}~\bibnamefont
  {Wang}},\ }\href {\doibase 10.1103/PhysRevD.79.055024} {\bibfield  {journal}
  {\bibinfo  {journal} {Phys. Rev. D}\ }\textbf {\bibinfo {volume} {79}},\
  \bibinfo {pages} {055024} (\bibinfo {year} {2009})},\ \Eprint
  {http://arxiv.org/abs/0811.3957} {arXiv:0811.3957 [hep-ph]} \BibitemShut
  {NoStop}%
\bibitem [{\citenamefont {Cirelli}\ \emph {et~al.}(2006)\citenamefont
  {Cirelli}, \citenamefont {Fornengo},\ and\ \citenamefont
  {Strumia}}]{Cirelli:2005uq}%
  \BibitemOpen
  \bibfield  {author} {\bibinfo {author} {\bibfnamefont {M.}~\bibnamefont
  {Cirelli}}, \bibinfo {author} {\bibfnamefont {N.}~\bibnamefont {Fornengo}}, \
  and\ \bibinfo {author} {\bibfnamefont {A.}~\bibnamefont {Strumia}},\ }\href
  {\doibase 10.1016/j.nuclphysb.2006.07.012} {\bibfield  {journal} {\bibinfo
  {journal} {Nucl. Phys. B}\ }\textbf {\bibinfo {volume} {753}},\ \bibinfo
  {pages} {178} (\bibinfo {year} {2006})},\ \Eprint
  {http://arxiv.org/abs/hep-ph/0512090} {arXiv:hep-ph/0512090} \BibitemShut
  {NoStop}%
\bibitem [{\citenamefont {Bell}\ \emph {et~al.}(2020)\citenamefont {Bell},
  \citenamefont {Dolan}, \citenamefont {Friedrich}, \citenamefont
  {Ramsey-Musolf},\ and\ \citenamefont {Volkas}}]{Bell:2020gug}%
  \BibitemOpen
  \bibfield  {author} {\bibinfo {author} {\bibfnamefont {N.~F.}\ \bibnamefont
  {Bell}}, \bibinfo {author} {\bibfnamefont {M.~J.}\ \bibnamefont {Dolan}},
  \bibinfo {author} {\bibfnamefont {L.~S.}\ \bibnamefont {Friedrich}}, \bibinfo
  {author} {\bibfnamefont {M.~J.}\ \bibnamefont {Ramsey-Musolf}}, \ and\
  \bibinfo {author} {\bibfnamefont {R.~R.}\ \bibnamefont {Volkas}},\ }\href
  {\doibase 10.1007/JHEP05(2020)050} {\bibfield  {journal} {\bibinfo  {journal}
  {JHEP}\ }\textbf {\bibinfo {volume} {05}},\ \bibinfo {pages} {050} (\bibinfo
  {year} {2020})},\ \Eprint {http://arxiv.org/abs/2001.05335} {arXiv:2001.05335
  [hep-ph]} \BibitemShut {NoStop}%
\bibitem [{\citenamefont {Chiang}\ \emph {et~al.}(2021)\citenamefont {Chiang},
  \citenamefont {Cottin}, \citenamefont {Du}, \citenamefont {Fuyuto},\ and\
  \citenamefont {Ramsey-Musolf}}]{Chiang:2020rcv}%
  \BibitemOpen
  \bibfield  {author} {\bibinfo {author} {\bibfnamefont {C.-W.}\ \bibnamefont
  {Chiang}}, \bibinfo {author} {\bibfnamefont {G.}~\bibnamefont {Cottin}},
  \bibinfo {author} {\bibfnamefont {Y.}~\bibnamefont {Du}}, \bibinfo {author}
  {\bibfnamefont {K.}~\bibnamefont {Fuyuto}}, \ and\ \bibinfo {author}
  {\bibfnamefont {M.~J.}\ \bibnamefont {Ramsey-Musolf}},\ }\href {\doibase
  10.1007/JHEP01(2021)198} {\bibfield  {journal} {\bibinfo  {journal} {JHEP}\
  }\textbf {\bibinfo {volume} {01}},\ \bibinfo {pages} {198} (\bibinfo {year}
  {2021})},\ \Eprint {http://arxiv.org/abs/2003.07867} {arXiv:2003.07867
  [hep-ph]} \BibitemShut {NoStop}%
\bibitem [{\citenamefont {Niemi}\ \emph {et~al.}(2019)\citenamefont {Niemi},
  \citenamefont {Patel}, \citenamefont {Ramsey-Musolf}, \citenamefont
  {Tenkanen},\ and\ \citenamefont {Weir}}]{Niemi:2018asa}%
  \BibitemOpen
  \bibfield  {author} {\bibinfo {author} {\bibfnamefont {L.}~\bibnamefont
  {Niemi}}, \bibinfo {author} {\bibfnamefont {H.~H.}\ \bibnamefont {Patel}},
  \bibinfo {author} {\bibfnamefont {M.~J.}\ \bibnamefont {Ramsey-Musolf}},
  \bibinfo {author} {\bibfnamefont {T.~V.~I.}\ \bibnamefont {Tenkanen}}, \ and\
  \bibinfo {author} {\bibfnamefont {D.~J.}\ \bibnamefont {Weir}},\ }\href
  {\doibase 10.1103/PhysRevD.100.035002} {\bibfield  {journal} {\bibinfo
  {journal} {Phys. Rev.}\ }\textbf {\bibinfo {volume} {D100}},\ \bibinfo
  {pages} {035002} (\bibinfo {year} {2019})},\ \Eprint
  {http://arxiv.org/abs/1802.10500} {arXiv:1802.10500 [hep-ph]} \BibitemShut
  {NoStop}%
%%CITATION = ARXIV:1802.10500;%%
\bibitem [{\citenamefont {Georgi}\ and\ \citenamefont
  {Glashow}(1972)}]{Georgi:1972cj}%
  \BibitemOpen
  \bibfield  {author} {\bibinfo {author} {\bibfnamefont {H.}~\bibnamefont
  {Georgi}}\ and\ \bibinfo {author} {\bibfnamefont {S.~L.}\ \bibnamefont
  {Glashow}},\ }\href {\doibase 10.1103/PhysRevLett.28.1494} {\bibfield
  {journal} {\bibinfo  {journal} {Phys. Rev. Lett.}\ }\textbf {\bibinfo
  {volume} {28}},\ \bibinfo {pages} {1494} (\bibinfo {year}
  {1972})}\BibitemShut {NoStop}%
%%CITATION = PRLTA,28,1494;%%
\bibitem [{\citenamefont {Shnir}(2005)}]{Balian:2005joa}%
  \BibitemOpen
  \bibfield  {author} {\bibinfo {author} {\bibfnamefont {Y.~M.}\ \bibnamefont
  {Shnir}},\ }\href {\doibase 10.1007/3-540-29082-6} {\emph {\bibinfo {title}
  {{Magnetic Monopoles}}}},\ Text and Monographs in Physics\ (\bibinfo
  {publisher} {Springer},\ \bibinfo {address} {Berlin/Heidelberg},\ \bibinfo
  {year} {2005})\BibitemShut {NoStop}%
\bibitem [{\citenamefont {'t~Hooft}(1974)}]{tHooft:1974kcl}%
  \BibitemOpen
  \bibfield  {author} {\bibinfo {author} {\bibfnamefont {G.}~\bibnamefont
  {'t~Hooft}},\ }\href {\doibase 10.1016/0550-3213(74)90486-6} {\bibfield
  {journal} {\bibinfo  {journal} {Nucl. Phys. B}\ }\textbf {\bibinfo {volume}
  {79}},\ \bibinfo {pages} {276} (\bibinfo {year} {1974})}\BibitemShut
  {NoStop}%
\bibitem [{\citenamefont {Polyakov}(1974)}]{Polyakov:1974ek}%
  \BibitemOpen
  \bibfield  {author} {\bibinfo {author} {\bibfnamefont {A.~M.}\ \bibnamefont
  {Polyakov}},\ }\href@noop {} {\bibfield  {journal} {\bibinfo  {journal} {JETP
  Lett.}\ }\textbf {\bibinfo {volume} {20}},\ \bibinfo {pages} {194} (\bibinfo
  {year} {1974})}\BibitemShut {NoStop}%
\bibitem [{\citenamefont {Rajantie}(2003)}]{Rajantie:2002dw}%
  \BibitemOpen
  \bibfield  {author} {\bibinfo {author} {\bibfnamefont {A.}~\bibnamefont
  {Rajantie}},\ }\href {\doibase 10.1103/PhysRevD.68.021301} {\bibfield
  {journal} {\bibinfo  {journal} {Phys. Rev. D}\ }\textbf {\bibinfo {volume}
  {68}},\ \bibinfo {pages} {021301} (\bibinfo {year} {2003})},\ \Eprint
  {http://arxiv.org/abs/hep-ph/0212130} {arXiv:hep-ph/0212130} \BibitemShut
  {NoStop}%
\bibitem [{\citenamefont {Kajantie}\ \emph
  {et~al.}(1997{\natexlab{a}})\citenamefont {Kajantie}, \citenamefont {Laine},
  \citenamefont {Rummukainen},\ and\ \citenamefont
  {Shaposhnikov}}]{Kajantie:1997tt}%
  \BibitemOpen
  \bibfield  {author} {\bibinfo {author} {\bibfnamefont {K.}~\bibnamefont
  {Kajantie}}, \bibinfo {author} {\bibfnamefont {M.}~\bibnamefont {Laine}},
  \bibinfo {author} {\bibfnamefont {K.}~\bibnamefont {Rummukainen}}, \ and\
  \bibinfo {author} {\bibfnamefont {M.~E.}\ \bibnamefont {Shaposhnikov}},\
  }\href {\doibase 10.1016/S0550-3213(97)00425-2} {\bibfield  {journal}
  {\bibinfo  {journal} {Nucl. Phys.}\ }\textbf {\bibinfo {volume} {B503}},\
  \bibinfo {pages} {357} (\bibinfo {year} {1997}{\natexlab{a}})},\ \Eprint
  {http://arxiv.org/abs/hep-ph/9704416} {arXiv:hep-ph/9704416 [hep-ph]}
  \BibitemShut {NoStop}%
%%CITATION = HEP-PH/9704416;%%
\bibitem [{\citenamefont {Kajantie}\ \emph
  {et~al.}(1997{\natexlab{b}})\citenamefont {Kajantie}, \citenamefont {Laine},
  \citenamefont {Rummukainen},\ and\ \citenamefont
  {Shaposhnikov}}]{Kajantie:1996qd}%
  \BibitemOpen
  \bibfield  {author} {\bibinfo {author} {\bibfnamefont {K.}~\bibnamefont
  {Kajantie}}, \bibinfo {author} {\bibfnamefont {M.}~\bibnamefont {Laine}},
  \bibinfo {author} {\bibfnamefont {K.}~\bibnamefont {Rummukainen}}, \ and\
  \bibinfo {author} {\bibfnamefont {M.~E.}\ \bibnamefont {Shaposhnikov}},\
  }\href {\doibase 10.1016/S0550-3213(97)00164-8} {\bibfield  {journal}
  {\bibinfo  {journal} {Nucl. Phys.}\ }\textbf {\bibinfo {volume} {B493}},\
  \bibinfo {pages} {413} (\bibinfo {year} {1997}{\natexlab{b}})},\ \Eprint
  {http://arxiv.org/abs/hep-lat/9612006} {arXiv:hep-lat/9612006 [hep-lat]}
  \BibitemShut {NoStop}%
%%CITATION = HEP-LAT/9612006;%%
\bibitem [{\citenamefont {Arnold}\ and\ \citenamefont
  {Espinosa}(1993)}]{Arnold:1992rz}%
  \BibitemOpen
  \bibfield  {author} {\bibinfo {author} {\bibfnamefont {P.~B.}\ \bibnamefont
  {Arnold}}\ and\ \bibinfo {author} {\bibfnamefont {O.}~\bibnamefont
  {Espinosa}},\ }\href {\doibase 10.1103/physrevd.50.6662.2,
  10.1103/PhysRevD.47.3546} {\bibfield  {journal} {\bibinfo  {journal} {Phys.
  Rev.}\ }\textbf {\bibinfo {volume} {D47}},\ \bibinfo {pages} {3546} (\bibinfo
  {year} {1993})},\ \bibinfo {note} {[Erratum: Phys. Rev.D50,6662(1994)]},\
  \Eprint {http://arxiv.org/abs/hep-ph/9212235} {arXiv:hep-ph/9212235 [hep-ph]}
  \BibitemShut {NoStop}%
%%CITATION = HEP-PH/9212235;%%
\bibitem [{\citenamefont {Laine}(1995{\natexlab{a}})}]{Laine:1994zq}%
  \BibitemOpen
  \bibfield  {author} {\bibinfo {author} {\bibfnamefont {M.}~\bibnamefont
  {Laine}},\ }\href {\doibase 10.1103/PhysRevD.51.4525} {\bibfield  {journal}
  {\bibinfo  {journal} {Phys. Rev.}\ }\textbf {\bibinfo {volume} {D51}},\
  \bibinfo {pages} {4525} (\bibinfo {year} {1995}{\natexlab{a}})},\ \Eprint
  {http://arxiv.org/abs/hep-ph/9411252} {arXiv:hep-ph/9411252 [hep-ph]}
  \BibitemShut {NoStop}%
%%CITATION = HEP-PH/9411252;%%
\bibitem [{\citenamefont {Patel}\ and\ \citenamefont
  {Ramsey-Musolf}(2011)}]{Patel:2011th}%
  \BibitemOpen
  \bibfield  {author} {\bibinfo {author} {\bibfnamefont {H.~H.}\ \bibnamefont
  {Patel}}\ and\ \bibinfo {author} {\bibfnamefont {M.~J.}\ \bibnamefont
  {Ramsey-Musolf}},\ }\href {\doibase 10.1007/JHEP07(2011)029} {\bibfield
  {journal} {\bibinfo  {journal} {JHEP}\ }\textbf {\bibinfo {volume} {07}},\
  \bibinfo {pages} {029} (\bibinfo {year} {2011})},\ \Eprint
  {http://arxiv.org/abs/1101.4665} {arXiv:1101.4665 [hep-ph]} \BibitemShut
  {NoStop}%
%%CITATION = ARXIV:1101.4665;%%
\bibitem [{\citenamefont {Kajantie}\ \emph
  {et~al.}(1996{\natexlab{c}})\citenamefont {Kajantie}, \citenamefont {Laine},
  \citenamefont {Rummukainen},\ and\ \citenamefont
  {Shaposhnikov}}]{Kajantie:1995kf}%
  \BibitemOpen
  \bibfield  {author} {\bibinfo {author} {\bibfnamefont {K.}~\bibnamefont
  {Kajantie}}, \bibinfo {author} {\bibfnamefont {M.}~\bibnamefont {Laine}},
  \bibinfo {author} {\bibfnamefont {K.}~\bibnamefont {Rummukainen}}, \ and\
  \bibinfo {author} {\bibfnamefont {M.~E.}\ \bibnamefont {Shaposhnikov}},\
  }\href {\doibase 10.1016/0550-3213(96)00052-1} {\bibfield  {journal}
  {\bibinfo  {journal} {Nucl. Phys.}\ }\textbf {\bibinfo {volume} {B466}},\
  \bibinfo {pages} {189} (\bibinfo {year} {1996}{\natexlab{c}})},\ \Eprint
  {http://arxiv.org/abs/hep-lat/9510020} {arXiv:hep-lat/9510020 [hep-lat]}
  \BibitemShut {NoStop}%
%%CITATION = HEP-LAT/9510020;%%
\bibitem [{\citenamefont {Laine}(1995{\natexlab{b}})}]{Laine:1995np}%
  \BibitemOpen
  \bibfield  {author} {\bibinfo {author} {\bibfnamefont {M.}~\bibnamefont
  {Laine}},\ }\href {\doibase 10.1016/0550-3213(95)00356-W} {\bibfield
  {journal} {\bibinfo  {journal} {Nucl. Phys.}\ }\textbf {\bibinfo {volume}
  {B451}},\ \bibinfo {pages} {484} (\bibinfo {year} {1995}{\natexlab{b}})},\
  \Eprint {http://arxiv.org/abs/hep-lat/9504001} {arXiv:hep-lat/9504001
  [hep-lat]} \BibitemShut {NoStop}%
%%CITATION = HEP-LAT/9504001;%%
\bibitem [{\citenamefont {Moore}\ and\ \citenamefont
  {Rummukainen}(2001)}]{Moore:2000jw}%
  \BibitemOpen
  \bibfield  {author} {\bibinfo {author} {\bibfnamefont {G.~D.}\ \bibnamefont
  {Moore}}\ and\ \bibinfo {author} {\bibfnamefont {K.}~\bibnamefont
  {Rummukainen}},\ }\href {\doibase 10.1103/PhysRevD.63.045002} {\bibfield
  {journal} {\bibinfo  {journal} {Phys. Rev.}\ }\textbf {\bibinfo {volume}
  {D63}},\ \bibinfo {pages} {045002} (\bibinfo {year} {2001})},\ \Eprint
  {http://arxiv.org/abs/hep-ph/0009132} {arXiv:hep-ph/0009132 [hep-ph]}
  \BibitemShut {NoStop}%
%%CITATION = HEP-PH/0009132;%%
\bibitem [{\citenamefont {Berg}\ and\ \citenamefont
  {Neuhaus}(1991)}]{Berg:1991cf}%
  \BibitemOpen
  \bibfield  {author} {\bibinfo {author} {\bibfnamefont {B.~A.}\ \bibnamefont
  {Berg}}\ and\ \bibinfo {author} {\bibfnamefont {T.}~\bibnamefont {Neuhaus}},\
  }\href {\doibase 10.1016/0370-2693(91)91256-U} {\bibfield  {journal}
  {\bibinfo  {journal} {Phys. Lett.}\ }\textbf {\bibinfo {volume} {B267}},\
  \bibinfo {pages} {249} (\bibinfo {year} {1991})}\BibitemShut {NoStop}%
%%CITATION = PHLTA,B267,249;%%
\bibitem [{\citenamefont {Ferrenberg}\ and\ \citenamefont
  {Swendsen}(1988)}]{Ferrenberg:1988yz}%
  \BibitemOpen
  \bibfield  {author} {\bibinfo {author} {\bibfnamefont {A.~M.}\ \bibnamefont
  {Ferrenberg}}\ and\ \bibinfo {author} {\bibfnamefont {R.~H.}\ \bibnamefont
  {Swendsen}},\ }\href {\doibase 10.1103/PhysRevLett.61.2635} {\bibfield
  {journal} {\bibinfo  {journal} {Phys. Rev. Lett.}\ }\textbf {\bibinfo
  {volume} {61}},\ \bibinfo {pages} {2635} (\bibinfo {year}
  {1988})}\BibitemShut {NoStop}%
%%CITATION = PRLTA,61,2635;%%
\bibitem [{\citenamefont {Laine}\ and\ \citenamefont
  {Rummukainen}(1998)}]{Laine:1998qk}%
  \BibitemOpen
  \bibfield  {author} {\bibinfo {author} {\bibfnamefont {M.}~\bibnamefont
  {Laine}}\ and\ \bibinfo {author} {\bibfnamefont {K.}~\bibnamefont
  {Rummukainen}},\ }\href {\doibase 10.1016/S0550-3213(98)00530-6} {\bibfield
  {journal} {\bibinfo  {journal} {Nucl. Phys.}\ }\textbf {\bibinfo {volume}
  {B535}},\ \bibinfo {pages} {423} (\bibinfo {year} {1998})},\ \Eprint
  {http://arxiv.org/abs/hep-lat/9804019} {arXiv:hep-lat/9804019 [hep-lat]}
  \BibitemShut {NoStop}%
%%CITATION = HEP-LAT/9804019;%%
\bibitem [{\citenamefont {Wang}\ and\ \citenamefont
  {Landau}(2001)}]{Wang:2000fzi}%
  \BibitemOpen
  \bibfield  {author} {\bibinfo {author} {\bibfnamefont {F.}~\bibnamefont
  {Wang}}\ and\ \bibinfo {author} {\bibfnamefont {D.~P.}\ \bibnamefont
  {Landau}},\ }\href {\doibase 10.1103/PhysRevLett.86.2050} {\bibfield
  {journal} {\bibinfo  {journal} {Phys. Rev. Lett.}\ }\textbf {\bibinfo
  {volume} {86}},\ \bibinfo {pages} {2050} (\bibinfo {year} {2001})},\ \Eprint
  {http://arxiv.org/abs/cond-mat/0011174} {arXiv:cond-mat/0011174
  [cond-mat.stat-mech]} \BibitemShut {NoStop}%
%%CITATION = COND-MAT/0011174;%%
\bibitem [{\citenamefont {Laine}\ and\ \citenamefont
  {Rummukainen}(2001)}]{Laine:2000rm}%
  \BibitemOpen
  \bibfield  {author} {\bibinfo {author} {\bibfnamefont {M.}~\bibnamefont
  {Laine}}\ and\ \bibinfo {author} {\bibfnamefont {K.}~\bibnamefont
  {Rummukainen}},\ }\href {\doibase 10.1016/S0550-3213(00)00736-7} {\bibfield
  {journal} {\bibinfo  {journal} {Nucl. Phys.}\ }\textbf {\bibinfo {volume}
  {B597}},\ \bibinfo {pages} {23} (\bibinfo {year} {2001})},\ \Eprint
  {http://arxiv.org/abs/hep-lat/0009025} {arXiv:hep-lat/0009025 [hep-lat]}
  \BibitemShut {NoStop}%
%%CITATION = HEP-LAT/0009025;%%
\bibitem [{\citenamefont {Laine}\ \emph {et~al.}(2013)\citenamefont {Laine},
  \citenamefont {Nardini},\ and\ \citenamefont {Rummukainen}}]{Laine:2012jy}%
  \BibitemOpen
  \bibfield  {author} {\bibinfo {author} {\bibfnamefont {M.}~\bibnamefont
  {Laine}}, \bibinfo {author} {\bibfnamefont {G.}~\bibnamefont {Nardini}}, \
  and\ \bibinfo {author} {\bibfnamefont {K.}~\bibnamefont {Rummukainen}},\
  }\href {\doibase 10.1088/1475-7516/2013/01/011} {\bibfield  {journal}
  {\bibinfo  {journal} {JCAP}\ }\textbf {\bibinfo {volume} {1301}},\ \bibinfo
  {pages} {011} (\bibinfo {year} {2013})},\ \Eprint
  {http://arxiv.org/abs/1211.7344} {arXiv:1211.7344 [hep-ph]} \BibitemShut
  {NoStop}%
%%CITATION = ARXIV:1211.7344;%%
\bibitem [{\citenamefont {Kennedy}\ and\ \citenamefont
  {Pendleton}(1985)}]{Kennedy:1985nu}%
  \BibitemOpen
  \bibfield  {author} {\bibinfo {author} {\bibfnamefont {A.~D.}\ \bibnamefont
  {Kennedy}}\ and\ \bibinfo {author} {\bibfnamefont {B.~J.}\ \bibnamefont
  {Pendleton}},\ }\href {\doibase 10.1016/0370-2693(85)91632-6} {\bibfield
  {journal} {\bibinfo  {journal} {Phys. Lett.}\ }\textbf {\bibinfo {volume}
  {156B}},\ \bibinfo {pages} {393} (\bibinfo {year} {1985})}\BibitemShut
  {NoStop}%
%%CITATION = PHLTA,156B,393;%%
\bibitem [{\citenamefont {Farakos}\ \emph {et~al.}(1995)\citenamefont
  {Farakos}, \citenamefont {Kajantie}, \citenamefont {Rummukainen},\ and\
  \citenamefont {Shaposhnikov}}]{Farakos:1994xh}%
  \BibitemOpen
  \bibfield  {author} {\bibinfo {author} {\bibfnamefont {K.}~\bibnamefont
  {Farakos}}, \bibinfo {author} {\bibfnamefont {K.}~\bibnamefont {Kajantie}},
  \bibinfo {author} {\bibfnamefont {K.}~\bibnamefont {Rummukainen}}, \ and\
  \bibinfo {author} {\bibfnamefont {M.~E.}\ \bibnamefont {Shaposhnikov}},\
  }\href {\doibase 10.1016/0550-3213(95)80129-4} {\bibfield  {journal}
  {\bibinfo  {journal} {Nucl. Phys.}\ }\textbf {\bibinfo {volume} {B442}},\
  \bibinfo {pages} {317} (\bibinfo {year} {1995})},\ \Eprint
  {http://arxiv.org/abs/hep-lat/9412091} {arXiv:hep-lat/9412091 [hep-lat]}
  \BibitemShut {NoStop}%
%%CITATION = HEP-LAT/9412091;%%
\bibitem [{\citenamefont {Davis}\ \emph {et~al.}(2002)\citenamefont {Davis},
  \citenamefont {Hart}, \citenamefont {Kibble},\ and\ \citenamefont
  {Rajantie}}]{Davis:2001mg}%
  \BibitemOpen
  \bibfield  {author} {\bibinfo {author} {\bibfnamefont {A.}~\bibnamefont
  {Davis}}, \bibinfo {author} {\bibfnamefont {A.}~\bibnamefont {Hart}},
  \bibinfo {author} {\bibfnamefont {T.}~\bibnamefont {Kibble}}, \ and\ \bibinfo
  {author} {\bibfnamefont {A.}~\bibnamefont {Rajantie}},\ }\href {\doibase
  10.1103/PhysRevD.65.125008} {\bibfield  {journal} {\bibinfo  {journal} {Phys.
  Rev. D}\ }\textbf {\bibinfo {volume} {65}},\ \bibinfo {pages} {125008}
  (\bibinfo {year} {2002})},\ \Eprint {http://arxiv.org/abs/hep-lat/0110154}
  {arXiv:hep-lat/0110154} \BibitemShut {NoStop}%
\bibitem [{\citenamefont {D'Onofrio}\ \emph {et~al.}(2014)\citenamefont
  {D'Onofrio}, \citenamefont {Rummukainen},\ and\ \citenamefont
  {Tranberg}}]{DOnofrio:2014rug}%
  \BibitemOpen
  \bibfield  {author} {\bibinfo {author} {\bibfnamefont {M.}~\bibnamefont
  {D'Onofrio}}, \bibinfo {author} {\bibfnamefont {K.}~\bibnamefont
  {Rummukainen}}, \ and\ \bibinfo {author} {\bibfnamefont {A.}~\bibnamefont
  {Tranberg}},\ }\href {\doibase 10.1103/PhysRevLett.113.141602} {\bibfield
  {journal} {\bibinfo  {journal} {Phys.\ Rev.\ Lett.}\ }\textbf {\bibinfo
  {volume} {113}},\ \bibinfo {pages} {141602} (\bibinfo {year} {2014})},\
  \Eprint {http://arxiv.org/abs/1404.3565} {arXiv:1404.3565 [hep-ph]}
  \BibitemShut {NoStop}%
\bibitem [{\citenamefont {Rubakov}(1981)}]{Rubakov:1981rg}%
  \BibitemOpen
  \bibfield  {author} {\bibinfo {author} {\bibfnamefont {V.}~\bibnamefont
  {Rubakov}},\ }\href@noop {} {\bibfield  {journal} {\bibinfo  {journal} {JETP
  Lett.}\ }\textbf {\bibinfo {volume} {33}},\ \bibinfo {pages} {644} (\bibinfo
  {year} {1981})}\BibitemShut {NoStop}%
\bibitem [{\citenamefont {Farakos}\ \emph {et~al.}(1994)\citenamefont
  {Farakos}, \citenamefont {Kajantie}, \citenamefont {Rummukainen},\ and\
  \citenamefont {Shaposhnikov}}]{Farakos:1994kx}%
  \BibitemOpen
  \bibfield  {author} {\bibinfo {author} {\bibfnamefont {K.}~\bibnamefont
  {Farakos}}, \bibinfo {author} {\bibfnamefont {K.}~\bibnamefont {Kajantie}},
  \bibinfo {author} {\bibfnamefont {K.}~\bibnamefont {Rummukainen}}, \ and\
  \bibinfo {author} {\bibfnamefont {M.~E.}\ \bibnamefont {Shaposhnikov}},\
  }\href {\doibase 10.1016/0550-3213(94)90173-2} {\bibfield  {journal}
  {\bibinfo  {journal} {Nucl. Phys.}\ }\textbf {\bibinfo {volume} {B425}},\
  \bibinfo {pages} {67} (\bibinfo {year} {1994})},\ \Eprint
  {http://arxiv.org/abs/hep-ph/9404201} {arXiv:hep-ph/9404201 [hep-ph]}
  \BibitemShut {NoStop}%
%%CITATION = HEP-PH/9404201;%%
\bibitem [{\citenamefont {Chetyrkin}\ and\ \citenamefont
  {Tkachov}(1981)}]{Chetyrkin:1981qh}%
  \BibitemOpen
  \bibfield  {author} {\bibinfo {author} {\bibfnamefont {K.~G.}\ \bibnamefont
  {Chetyrkin}}\ and\ \bibinfo {author} {\bibfnamefont {F.~V.}\ \bibnamefont
  {Tkachov}},\ }\href {\doibase 10.1016/0550-3213(81)90199-1} {\bibfield
  {journal} {\bibinfo  {journal} {Nucl. Phys.}\ }\textbf {\bibinfo {volume}
  {B192}},\ \bibinfo {pages} {159} (\bibinfo {year} {1981})}\BibitemShut
  {NoStop}%
%%CITATION = NUPHA,B192,159;%%
\bibitem [{\citenamefont {Laporta}(2000)}]{Laporta:2001dd}%
  \BibitemOpen
  \bibfield  {author} {\bibinfo {author} {\bibfnamefont {S.}~\bibnamefont
  {Laporta}},\ }\href {\doibase 10.1016/S0217-751X(00)00215-7,
  10.1142/S0217751X00002157} {\bibfield  {journal} {\bibinfo  {journal} {Int.
  J. Mod. Phys.}\ }\textbf {\bibinfo {volume} {A15}},\ \bibinfo {pages} {5087}
  (\bibinfo {year} {2000})},\ \Eprint {http://arxiv.org/abs/hep-ph/0102033}
  {arXiv:hep-ph/0102033 [hep-ph]} \BibitemShut {NoStop}%
%%CITATION = HEP-PH/0102033;%%
\bibitem [{\citenamefont {Nishimura}\ and\ \citenamefont
  {Schroder}(2012)}]{Nishimura:2012ee}%
  \BibitemOpen
  \bibfield  {author} {\bibinfo {author} {\bibfnamefont {M.}~\bibnamefont
  {Nishimura}}\ and\ \bibinfo {author} {\bibfnamefont {Y.}~\bibnamefont
  {Schroder}},\ }\href {\doibase 10.1007/JHEP09(2012)051} {\bibfield  {journal}
  {\bibinfo  {journal} {JHEP}\ }\textbf {\bibinfo {volume} {09}},\ \bibinfo
  {pages} {051} (\bibinfo {year} {2012})},\ \Eprint
  {http://arxiv.org/abs/1207.4042} {arXiv:1207.4042 [hep-ph]} \BibitemShut
  {NoStop}%
%%CITATION = ARXIV:1207.4042;%%
\end{thebibliography}%

%%%%%%%%%%%%%%%%%%%%%%%%%%%%%%%%%%%%%%%%%%%%%%%%%%%%%%

\end{document}